\documentclass[default,iicol]{sn-jnl}
\usepackage{commath}
\usepackage{multicol}
\usepackage{cleveref}
\crefname{fig}{Fig.}{Figs.}
\crefname{extendedfig}{Extended Data Fig.}{Extended Data Figs.}
\crefname{eq}{Eq.}{Eqs.}
\crefname{Methods}{Methods }{Methods Sections}

\usepackage[switch]{lineno}
\usepackage{titlesec}
\usepackage{soul}

\usepackage{dblfloatfix}

\jyear{2023}%
\theoremstyle{thmstyleone}%

\theoremstyle{thmstyletwo}%

\theoremstyle{thmstylethree}%
\begin{document}

\title[Article Title]{High-fidelity parametric beamsplitting with a parity-protected converter}

\author*[1,2]{\fnm{Yao} \sur{Lu}}\email{physics.lu@yale.edu}
\equalcont{These authors contributed equally to this work.}

\author*[1,2]{\fnm{Aniket} \sur{Maiti}}\email{aniket.maiti@yale.edu}
\equalcont{These authors contributed equally to this work.}

\author[1,2]{\fnm{John W. O.} \sur{Garmon}}
\author[1,2]{\fnm{Suhas} \sur{Ganjam}}
\author[1,2]{\fnm{Yaxing} \sur{Zhang}}
\author[1,2]{\fnm{Jahan} \sur{Claes}}
\author[1,2]{\fnm{Luigi} \sur{Frunzio}}
\author[1,2]{\fnm{Steven M.} \sur{Girvin}}
\author*[1,2]{\fnm{Robert J.} \sur{Schoelkopf}}\email{robert.schoelkopf@yale.edu}

\affil[1]{\orgdiv{Departments of Applied Physics and Physics}, \orgname{Yale University}, \orgaddress{\city{New Haven}, \postcode{06511}, \state{CT}, \country{USA}}}

\affil[2]{\orgdiv{Yale Quantum Institute}, \orgname{Yale University}, \orgaddress{\city{New Haven}, \postcode{06511}, \state{CT}, \country{USA}}}

\abstract{
Fast, high-fidelity operations between microwave resonators are an important tool for bosonic quantum computation and simulation with superconducting circuits. 
An attractive approach for implementing these operations is to couple these resonators via a nonlinear converter and actuate parametric processes with RF drives. 
It can be challenging to make these processes simultaneously fast and high fidelity, since this requires introducing strong drives without activating parasitic processes or introducing additional decoherence channels.
We show that in addition to a careful management of drive frequencies and the spectrum of environmental noise, leveraging the inbuilt symmetries of the converter Hamiltonian can suppress unwanted nonlinear interactions, preventing converter-induced decoherence.
We demonstrate these principles using a differentially-driven DC-SQUID as our converter, coupled to two high-Q microwave cavities.
Using this architecture, we engineer a highly-coherent beamsplitter and fast ($\sim$ 100 ns) swaps between the cavities, limited primarily by their intrinsic single-photon loss.
We characterize this beamsplitter in the cavities' joint single-photon subspace, and show that we can detect and post-select photon loss events to achieve a beamsplitter gate fidelity exceeding 99.98$\%$, which to our knowledge far surpasses the current state of the art.
}


\newcommand{\red}[1]{\textcolor{red}{#1}}
\newcommand{\newtext}[1]{\textcolor{purple}{#1}}

\maketitle

\section*{Introduction}

The precise manipulation of high-Q bosonic modes is vital to experimentally explore a wide range of phenomena in many-body physics and quantum information.
A crucial component in this endeavor is a programmable two-mode interaction, exemplified by the time-dependent beamsplitter Hamiltonian $\hat{\mathcal{H}}_\text{BS}/\hbar = g_\text{BS}(t) \left(e^{i\varphi_\text{BS}} \hat{a}^\dagger \hat{b} + e^{-i\varphi_\text{BS}} \hat{a} \hat{b}^\dagger\right)$.
This controlled photon-exchange coupling is a requirement for continuous variable quantum computation \cite{Gaussian_QI}, and has direct applications in bosonic simulations of hopping models and lattice gauge theories \cite{YaoMottInsulator,CreutzLadderExperiment,JensCQEDLattices,Roushan2016}.
It is particularly appealing to implement such a Hamiltonian in circuit-QED, a flexible platform offering readily available nonlinear control of high-Q modes in superconducting resonators \cite{ReagorStubCavity}.
These resonators primarily lose coherence through single-photon loss \cite{ECDControlPaper,SergeMushroomCavity}, a `noise-bias' that has been utilized in important demonstrations including bosonic error-correction \cite{ChuangBosonicCodes,CochraneCatCode,Leghtas4CatCode,JamesDualRailPaper} beyond break-even \cite{BeyondBreakeven2016,VladGKPBeyondBreakeven,BinomialBeyondBreakeven}.
However, the experimental implementation of a fast, high-fidelity beamsplitter that preserves the long lifetime of these resonators and does not introduce additional decoherence has remained a challenge.
If realized, this interaction would help construct logical entangling gates between qubits encoded in the resonators~\cite{Yvonne2019_Nature_eSWAP,TakaPaper,Joshi2021}, and enhance long-distance interactions through microwave quantum buses \cite{PhillipeDeterministicRemoteEntanglementPitchAndCatch,AxlineOnDemandStateTransfer}.

Within the circuit-QED framework, interactions between linear superconducting resonators can be implemented by coupling them via a Josephson junction-based nonlinear `converter'~\cite{JPC_Paper,SPA_1,JAMPA}.
Driving this converter with RF drives can actuate a `parametric beamsplitter'~\cite{AumentadoDualRail,SiroisStorageRetrieval2015,Yvonne2018_PRX_ProgrammableInterference,FPJA,ParametricCQEDPaper,HatridgeRouter} between modes that are widely separated in frequency space, offering large on-off ratios.
The amplitude of the drives sets the strength of the beamsplitter, so one could ideally improve both the speed and fidelity of beamsplitting-based gates by simply driving harder.
However, controlling the dynamics of strongly driven, dissipative, nonlinear systems can be challenging~\cite{KorotkovBeyondRWAPaper,Yaxing_PRA_BilinearModeCoupling}.
The wide-bandwidth of the Josephson nonlinearity can activate numerous parasitic processes that cause exchanges with modes lossier than the resonators, including with the converter itself, spoiling fidelity.
The converter can also be incoherently excited through a dressing of its natural decoherence in the presence of the drives, which can directly dephase the beamsplitter interaction and harm the noise-bias of the resonators.
These parasitic processes and drive-induced coupler excitations have been significant barriers in previous implementations~\cite{Yvonne2018_PRX_ProgrammableInterference,Yvonne2019_Nature_eSWAP} and suppressing them is crucial to engineering a clean, high-fidelity beamsplitter.

One approach that has been used to tame this nonlinearity is engineering multi-junction converters with useful symmetries~\cite{JPC_Paper,ATS_Paper}, that prevent a significant fraction of the nonlinear processes allowed in a single-junction circuit like the transmon~\cite{JensTransmonPaper}.
In this work we employ a similar approach, using the familiar circuit of the symmetric DC Superconducting QUantum Interference Device (SQUID), with careful engineering to make it compatible with a high-Q environment.
Making full use of this symmetry requires the SQUID to be driven in a purely differential manner, and we introduce an architecture for delivering this drive through an auxiliary `buffer' mode.

In this work, we use the differentially-driven SQUID (DDS) to perform a fast, highly coherent beamsplitter between two high-Q 3D superconducting cavity resonators.
We show that the coupler experiences almost no drive-induced excitation, independent of the beamsplitter rate.
The beamsplitter fidelity is then limited only by the cavities' single-photon decay, thus being highly compatible with existing bosonic encodings~\cite{ATS_Paper,Grimm2020,Puri2020}.
As a first demonstration, we characterize the pulsed operation of this beamsplitter with single photons in a microwave implementation of the dual-rail qubit subspace $\{\vert 0_a 1_b \rangle$, $ \vert1_a0_b\rangle \} $, where the beamsplitter interaction provides universal control.
We achieve an average gate fidelity of $99.92\%$, which on detecting single-photon loss events is boosted to $99.98\%$. This paves the way for quantum computing architectures based on erasure-limited dual-rail qubits \cite{DivincenzoDualRail,AumentadoDualRail,JamesDualRailPaper}, and for cleaner parametric processes in general.

\begin{figure*}[ht]
\centering
\includegraphics[width=0.95\textwidth]{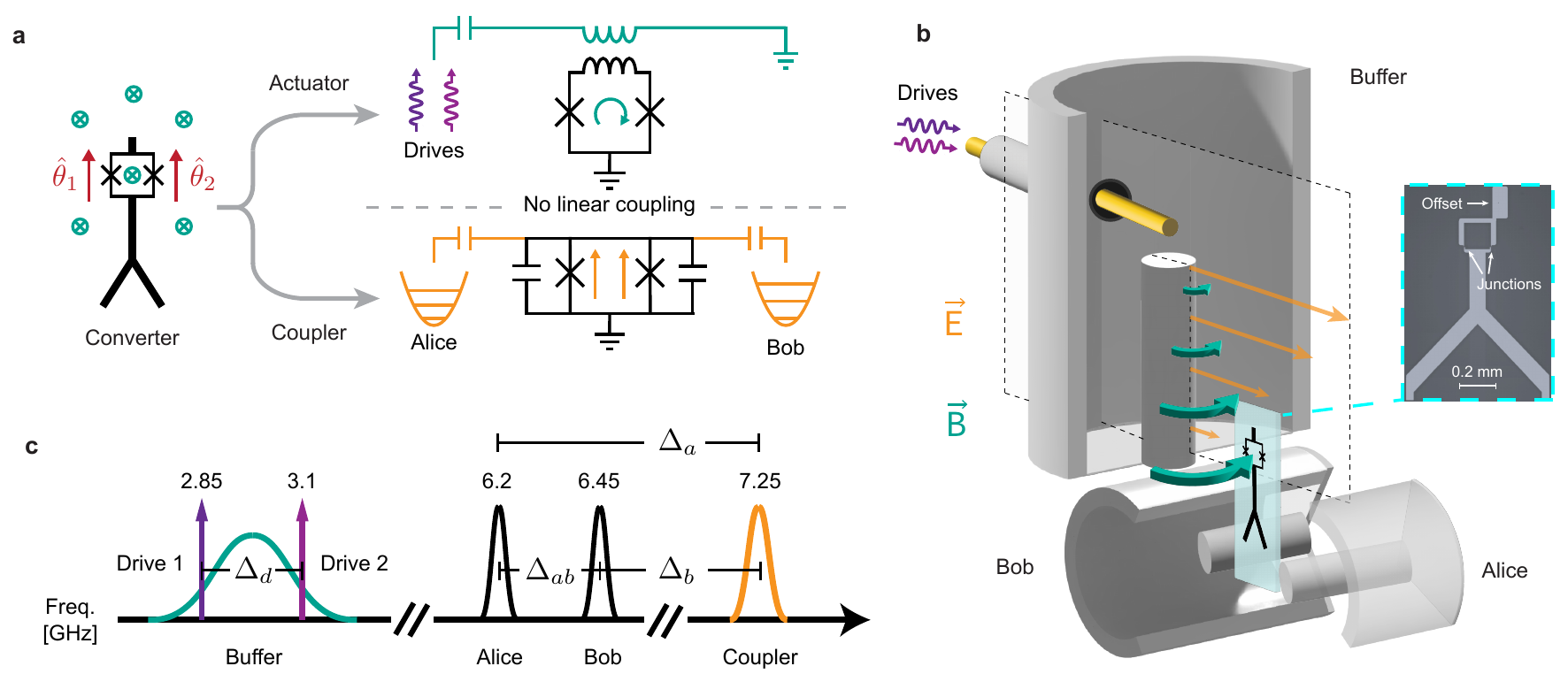}
\caption{
\textbf{The differentially driven SQUID as a parity protected converter.}
\textbf{a}, The symmetric DC-SQUID contains two orthogonal modes \cite{SQUIDModesLecocq2011,KamalSQUIDModes}, the common mode (coupler) and the differential mode (actuator). We selectively couple the former to two bosonic modes and the latter to the drives to take advantage of the natural symmetries of the Hamiltonian in \cref{eq:diff}.
\textbf{b}, Implementing the purely differential drive through a 3D buffer post-cavity (figure is exaggerated for illustrative purposes).
The natural separation in electric and magnetic fields in the $\lambda/4$ mode is used to purely drive the actuator, without exciting the coupler.
The sensitive quantum information is stored in two high-Q $\lambda/4$ post-cavities (Alice and Bob) that participate in the coupler, enabling parametric beamsplitting between them.
The inset shows an optical micrograph of the SQUID device, displaying the purposely offset antenna pad that counters residual drive-asymmetry.
\textbf{c}, Frequency stack for relevant modes in the system.
The difference of the two drive frequencies ($\Delta_d$) is fixed to be equal to the cavity detuning ($\Delta_{ab}$) for resonant beamsplitting.
The drives are placed symmetrically around the buffer mode resonance, which is engineered to be far-red detuned from the coupler frequency.
}\label[fig]{fig1}
\end{figure*}

\section*{Results}
\subsection*{Engineering a differentially-driven SQUID}
We now outline the key insight of properly leveraging the symmetry of the SQUID for a cleaner beamsplitter, and describe an architecture that accomplishes it.
The SQUID has two orthogonal modes, a common `coupler' mode with symmetric Josephson phases across its two junctions, and a differential `actuator' mode with anti-symmetric phases.
Fully utilizing the symmetry of these modes requires a precisely-engineered selective coupling of the coupler and actuator to the resonators and the drive respectively (\cref{fig1}a), at the sweet-spot of zero DC flux.
This is different from a conventional flux-driven SQUID \cite{OliverJPA,FirstOrderSideband,MckayCoupler,YaoUniversalStabilization}, since extra care must be taken to ensure that the drive does not excite the common mode.
This drive must also be introduced in a minimally invasive manner that does not spoil the lifetime of either the resonators or the coupler.

The Hamiltonian of such a differentially-driven SQUID is (Supplementary Note 1):
\begin{linenomath*}
    \begin{equation}
        \hat{\mathcal{H}}_\text{DDS} = 4 E_C \hat{n}_c^2 - E_J \underbrace{\cos(\phi_d)\cos(\hat{\theta}_c)}_{\text{Even Parity}},
        \label[eq]{eq:diff}
    \end{equation}
\end{linenomath*}
where $\hat{\theta}_c = \left(\hat{\theta}_1+\hat{\theta}_2\right)/2$ and $n_c$ are the conjugate phase and charge variables describing the coupler mode, and $\phi_d = \left<\hat{\theta}_1-\hat{\theta}_2\right>/2$ is the classical response of the driven actuator, for junction phases $\hat{\theta}_{1,2}$.
$E_J$ and $E_C$ are the total Josephson and charging energies respectively.
The advantage of engineering this Hamiltonian is two-fold.
First, the Hamiltonian is protected from processes that involve an odd number of mode quanta, while preserving the strength of desired quadratic process that provides the beamsplitting interaction.
This `parity protection' forbids up to half the parasitic coherent processes allowed by a single-junction converter, including the swapping of sensitive information into the coupler that limited previous implementations~\cite{Yvonne2018_PRX_ProgrammableInterference}.
Second, driving through the orthogonal actuator port provides flexibility in choosing the drive frequencies, which we can utilize to drive far away from non-protected processes that could otherwise limit fidelity.
Combining these two advantages has the potential to allow the suppression of not only unwanted coherent processes, but also incoherent coupler excitations \cite{Yaxing_PRA_BilinearModeCoupling,DykmanHeating} due to coupler decoherence dressed by the drives (see Supplementary Note 2 for a detailed description).

Implementing the symmetric Hamiltonian in \cref{eq:diff} amounts to fabricating a SQUID with symmetric junctions, calibrating it to zero DC flux, and delivering a purely differential drive.
The first two conditions are possible to optimize in fabrication or calibrate out respectively, and have been described in Supplementary Note 7.
Engineering the differential drive, however, necessitates the delivery high-frequency flux in a superconducting package, while simultaneously eliminating stray electric fields that couple to the common mode of the SQUID and accounting for spatial gradients in the time-dependent flux (see Methods ``Fine-tuning a differential drive" for more details).

We simultaneously fulfill these constraints using the $\lambda/4$ mode of a stub cavity (dubbed the `buffer mode'), integrated into the same monolithic package that hosts the high-Q storage modes (\cref{fig1}b).
Once driven, the buffer mode provides an oscillating magnetic field that serves as the flux drive penetrating the SQUID loop.
We place the SQUID at the base of the buffer mode, which functions as a virtual ground, minimizing the common mode drive due to the driven electric field while maximizing the flux penetrating the SQUID loop.
The coupling to any remnant electric field is further suppressed by orienting the SQUID's electric dipole moment to be perpendicular to the driven field.
In addition to this minimization of the common-mode drive, driving through the buffer mode also has the benefit of imposing a finite bandwidth ($\sim250$ MHz) on the noise spectrum of the environment seen through the drive line.

Finally, the full elimination of the common mode drive is achieved by a deliberate offset of the SQUID's capacitive pad.
This counters the effects of flux gradients in our system, by engineering an additional flux-dependent electromotive force, which can be directly optimized using finite-element-method simulations (see Methods).
In experiment, imperfections in fabrication or placement of the SQUID chip might affect the achievable drive orthogonality.
However, we are able to directly estimate this orthogonality in our actual device by experimentally comparing the strength of an allowed process to one that should be forbidden by parity protection , finding a residual common drive on the order of $\vert \phi_c / \phi_d \vert \sim 1\%$ (see Methods ``Experimentally characterizing residual drive asymmetry").

\subsection*{Demonstrating a high-coherence
beamsplitter}\label{secBS}

We now present the full experimental realization (\cref{fig1}b) of a high-fidelity beamsplitter that strongly suppresses undesirable coupler heating.
Our construction consists of a high-purity aluminum package hosting three coaxial $\lambda/4$ stub cavity modes: the buffer mode and the two high-Q storage modes \cite{ReagorStubCavity}.
The storage modes are capacitively coupled to the Y-shaped antenna of the SQUID \cite{Ymon_Paper} and have a negligible mutual inductance to the SQUID loop, ensuring that they exclusively participate \cite{BBQPaper,EPRPaper} in the coupler mode:
\begin{linenomath*}
    \begin{equation}
        \hat{\theta}_c \approx \left( \frac{2E_C}{E_J} \right)^{\frac{1}{4}} \left(\frac{g_{a}}{\Delta_{a}} \; \hat{a} + \frac{g_{b}}{\Delta_{b}} \; \hat{b}  + \hat{c}\right) + \text{h.c.} \label[eq]{eq:participation}
    \end{equation}
\end{linenomath*}
Here $\hat{a}$, $\hat{b}$  are the ladder operators for our dressed storage modes, named Alice and Bob respectively, while $\hat{c}$ represents the dressed coupler.
The coupling strengths ($g_{a,b}$) and mode detunings ($\Delta_{a,b}$, \cref{fig1}c) between the storage modes and the coupler are chosen to be in the dispersive regime $\left( \frac{g_{a}}{\Delta_{a}}, \frac{g_{b}}{\Delta_{b}} \sim 0.1 \right)$.
Additionally, we can prepare and readout Fock states in Bob through a dispersively coupled ancilla transmon \cite{DispersiveUniversalControl} and a dedicated stripline readout resonator.
The SQUID coupler also includes a dedicated readout resonator and drive pin, for explicit characterization of frequency shifts and drive-induced excitation.
The measured device parameters are presented in Supplementary Note 12, and details of device fabrication can be found in Supplementary Note 11.

We activate and control the amplitude and phase of our beamsplitter with a bi-chromatic drive on the actuator: $\phi_{d_{1,2}}(t) = \vert \phi_{d_{1,2}} \vert \cos(\omega_{d_{1,2}}t + \varphi_{d_{1,2}})$.
When the difference in our drive frequencies ($\Delta_d$) is close to our cavity detuning ($\Delta_{ab}$), \cref{eq:diff} and \cref{eq:participation} combine to create a tunable beamsplitter Hamiltonian (see Methods for full derivation):
\begin{linenomath*}
    \begin{align}
        \hat{\mathcal{H}}_\text{BS}/\hbar =  \;&\Delta_\text{BS} \; \hat{a}^\dagger \hat{a} + g_\text{BS}\; (e^{i\varphi_\text{BS}} \hat{a}^\dagger \hat{b} + e^{-i \varphi_\text{BS}}\hat{a} \hat{b}^\dagger), \label[eq]{eq:beamsplitting}
        \\ \text{with \hspace{1em}}g_\text{BS} &\approx \;\dfrac{\omega_c}{2} \: \dfrac{g_{a}\;g_{b}}{\Delta_{a}\Delta_{b}} \;J_1\left(\vert \phi_{d_1} \vert \right)J_1\left(\vert \phi_{d_2} \vert \right), \label[eq]{eq:g_bs_value} \\
        \Delta_\text{BS} &= \Delta_{ab} - \Delta_d + \Delta_{\text{Z},ab} \label[eq]{eq:delta_bs_value}
    \end{align}
\end{linenomath*}
where $J_1(\vert \phi_{d_{1,2}}\vert)$ is the first-order Bessel function of the drive amplitudes, and $\varphi_\text{BS}$ is the beamsplitter phase, controlled by the relative phase of the drives.
The drives also induce an additional frequency offset, the relative AC-Zeeman shift of the cavities, $\Delta_{\text{Z},ab}$, and we experimentally find the amplitude-dependent condition $\Delta_\text{BS}=0$ to execute a resonant beamsplitter.

\begin{figure*}[ht]
\centering
\includegraphics[width=0.95\textwidth]{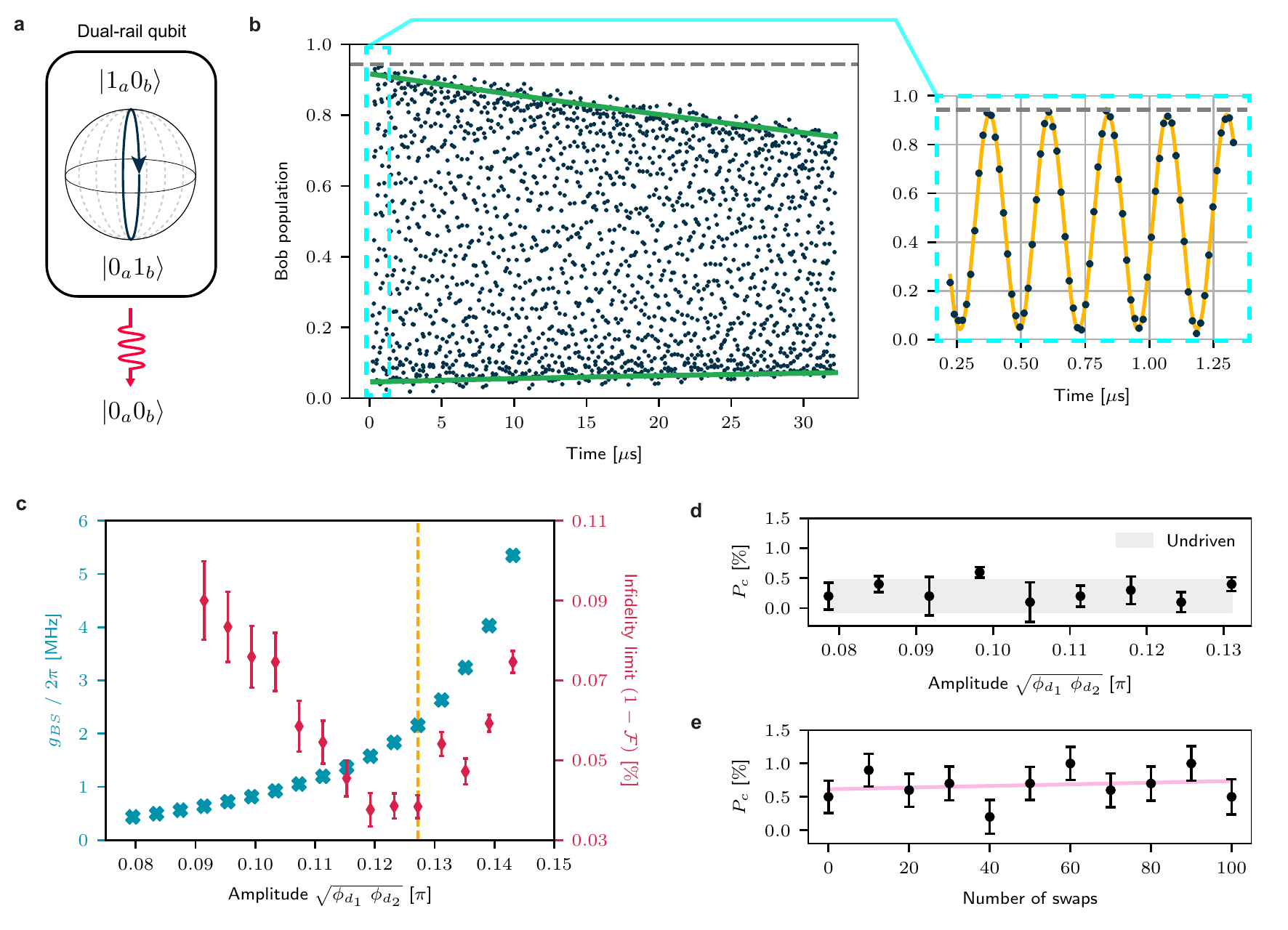}
\caption{ 
\textbf{Beamsplitting with the differentially-driven SQUID.}
\textbf{a}, Beamsplitting implements an effective driven Rabi evolution in the Bloch sphere of the dual-rail qubit formed by the single photon subspace Alice and Bob, where decay can be detected by monitoring the vacuum state.
\textbf{b}, Resonant evolution of a single-photon prepared in Bob.
The data is normalized for readout infidelity, and state preparation fidelity is shown as a dashed grey line (Supplementary Note 10).
The fast coherent oscillations (black dots) between the cavities are fitted to \cref{eq:Pbob} (green lines show envelope) to obtain the decay and dephasing time-scales.
The evolution for the first $1.5\; \mu$s is plotted separately to better illustrate the oscillations, and fit to a sinusoid to extract $g_{\text{BS}}$.
\textbf{c}, Sweeping both drive amplitudes simultaneously and repeating experiment \textbf{a} lets us quantify $g_\text{BS}$ (blue crosses), and the decoherence limit on beamsplitter infidelity (red diamonds) at various drive strengths.
We choose a drive strength with simultaneously low infidelity and high beamsplitter rate as our operating point (yellow dashed line).
\textbf{d}, The coupler's driven excitation ($P_c$) after evolving for 10 swaps is directly quantified through a dedicated on-chip readout mode.
We observe no monotonic correlation with respect to drive amplitude, and driven populations mostly remain within the range of the undriven population (grey region).
\textbf{e}, Coupler population as a function of number of swaps at the operational driving point.
The heating rate is nearly immeasurable, with a fitted (pink line) slope of $(1.2\pm2.4)\times 10^{-5}$ excitation per swap, which is within expectation for our natural thermal background ($\gamma_{c, \uparrow} \sim (3.3\textrm{ ms})^{-1}$).
The non-zero offset of the fit arises from preparation and readout infidelities.
Error bars in both \textbf{d} and \textbf{e} represent fit errors from the protocol described in \cite{GeerlingsRPM}.
}
\label[fig]{fig2}
\end{figure*}

We characterize our beamsplitter interaction using the joint single-photon subspace of our storage cavities, which forms a microwave implementation of a dual-rail qubit (\cref{fig2}a).
We initialize a single photon in Bob with a preparation fidelity of $\sim 94\%$, by displacing Bob to a coherent state ($\alpha_b=\sqrt{2}$) and using number-resolved measurements through the ancilla to  post-select the desired Fock state.
We then apply the resonant beamsplitter interaction for a range of times up to 32 $\mu$s (\cref{fig2}b), to estimate both the beamsplitting strength and the driven decoherence time of the dual-rail qubit.
The evolution of the time-dependent photon population in Bob follows (see Methods ``Decoherence in the dual-rail subspace" for derivation):
\begin{linenomath*}
    \begin{equation}
        P_\text{Bob} = \dfrac{1}{2}e^{-\kappa_1 t}\left(1 + e^{-\kappa_\varphi t} \; \cos{(2 g_\text{BS}t)}\right)
        \label[eq]{eq:Pbob}
    \end{equation}
\end{linenomath*}
Here, $\kappa_1$ is the mean of the driven cavities' single-photon decay rates, and is the effective rate of population leakage out of the dual-rail subspace into vacuum $ \vert 0_a 0_b \rangle $.
The dual-rail qubit may also experience dephasing at a rate $\kappa_{\varphi}$, which would drive this evolution towards an evenly mixed state within the qubit subspace.
Combining these lets us place a lower bound on the expected decoherence limit on the fidelity of a single beamsplitter operation (see Methods):
\begin{linenomath*}
    \begin{equation}
        \mathcal{F} \approx 1 - \frac{\pi}{4} \frac{\kappa_\text{BS}}{g_\text{BS}}, \hspace{1em} \kappa_\text{BS} = \kappa_1 + \frac{\kappa_\varphi}{2},
        \label[eq]{eq:fidbs}
    \end{equation}
\end{linenomath*}
which is a more accurate metric than the one used in \cite{Yvonne2018_PRX_ProgrammableInterference}.

We choose our operating point to maximize this expected fidelity $\mathcal{F}$ with respect to drive strength, which we characterize by analyzing sections of the resulting driven long-time evolution at various drive amplitudes (\cref{fig2}c).
At each amplitude, we extract the $g_\text{BS}$ and $\kappa_\text{BS}$ by fitting two short sections of the evolution (Methods) to \cref{eq:Pbob}.
We are able to obtain a maximum $g_\text{BS}/2\pi$ exceeding 5 MHz, with the effective decoherence-limited fidelity surpassing $99.9\%$ for a wide range of beamsplitting strengths.
We observe that upon increasing the drive strength, the coupler frequency shifts closer to the cavities (by $\sim200$ MHz), which may be accompanied by direct sideband interactions between the coupler and the cavities (Supplementary Note 5), leading to an increased hybridization between the modes.
This effect results in a faster-than-quadratic dependence of the beamsplitting strength on our drive amplitudes, but limits the fidelity of the beamsplitter at higher amplitudes due to the aggravated Purcell loss.

At the operating point, we fit the evolution in \cref{fig2}b to find $g_\text{BS}/2\pi  = 2.16 \pm 0.01 \textrm{MHz}$, $\kappa_1 = (197 \pm 8 \; \mu \mathrm{s})^{-1}$ and $\kappa_\varphi = (313 \pm 40 \; \mu \mathrm{s})^{-1}$.
Our effective $\kappa_\text{BS} =  (150 \pm 25 \; \mu \text{s})^{-1}$ places a decoherence-based upper bound on the fidelity of $\mathcal{F} = 99.96 \pm 0.01\%$,  which is almost two orders of magnitude better than the previous transmon-based implementation. Crucially, this fidelity is also  limited primarily by photon loss in the cavities, preserving their advantageous noise bias.

To directly quantify the suppression of drive-induced coupler excitation, we measure the coupler population as a function of drive amplitude after resonantly evolving for ten swaps.
With our coupler prepared in the ground state, we apply this pulse and measure the coupler's population through a protocol that is robust to readout infidelity \cite{GeerlingsRPM}.
We measure no correlated increase of its driven population as a function of drive amplitude, up to our measurement uncertainty of $\sim 0.2\%$ (\cref{fig2}d).
At the operating point, we explicitly quantify the increase in coupler excitation as a function of number of swaps (\cref{fig2}e).
We evolve the system up to 100 swaps and observe a total heating rate below $\sim 4\times 10^{-5}$ excitations per swap, which is consistent with the undriven heating rate of the coupler, implying no additional drive-induced heating.
This substantial suppression (three orders of magnitude better than transmon-based implementations \cite{Yvonne2018_PRX_ProgrammableInterference}) eliminates limitations placed by coupler-induced dephasing on the fidelity of the beamsplitter, allowing us to harness the long lifetimes and even longer dephasing times of the 3D cavities.

While this device was optimized for performance in the single-photon manifold, any drive-induced increase in the self-Kerr of the cavities would lead to coherent errors in higher-photon manifolds.
The sideband interaction resulting from our choice of mode frequencies exacerbates this effect, with up to 128 KHz of inherited Kerr at the operating point.
Numerous avenues exist to minimize this driven non-linearity if desired (see Supplementary Note 6), including choosing a slightly higher coupler frequency, arraying multiple SQUIDs, or dynamical Kerr cancellation \cite{YaxingKerrCancellation}.
If minimizing inherited self-Kerr is vital, such as in schemes utilizing large coherent states, then implementing these improvements or using an alternative scheme like Kerr-free three-wave mixing~\cite{SPA_1,ChapmanSNAILBeamsplitterPaper} may be required.

\begin{figure*}[ht]
\centering
\includegraphics[width=0.95\textwidth]{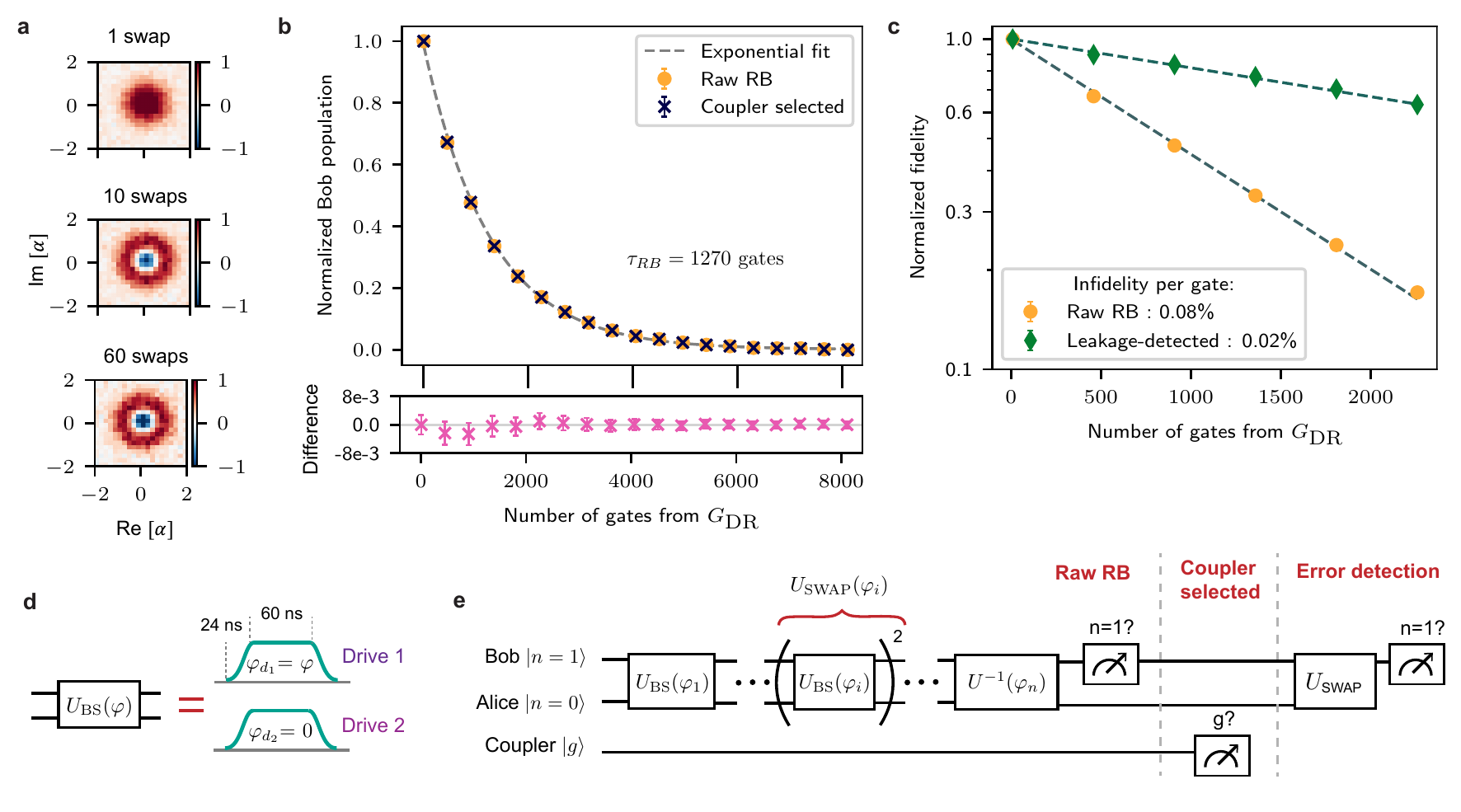}
\caption{
    \textbf{Randomized benchmarking with a calibrated beamsplitter pulse.}
    \textbf{a}, Wigner function of Bob after preparing $\vert 0_a 1_b \rangle$ and implementing 1, 10 and 60 calibrated swaps.
    Each swap is a combination of two identical beamplitter pulses.
    \textbf{b}, Probability of ending in the target state $\vert 0_a 1_b \rangle$ after executing the RB protocol with randomly selected gates from $G_{\textrm{DR}}$ (yellow circles).
    The curve is normalized to account for state preparation and readout imperfections (Supplementary Note 10), and is in good agreement with a single-exponential, with a decay constant of $\tau_{\text{RB}} = 1271 \pm 4$ gates.
    This `Raw RB' is practically indistinguishable from the sequences post-selected on the coupler ground state (black crosses), with the difference of the two curves shown below the main plot (pink crosses).
    \textbf{c}, Focusing on the first $2250$ gates, we use measurements of both cavities to post-select on sequences in which no photon loss event occurred (green diamonds).
    We compare these sequences to the  raw RB in \textbf{b} (yellow), showing an improvement in average gate infidelity from $0.078 \pm 0.001 \%$ to $0.020 \pm 0.001 \%$.
    \textbf{d}, The gate-sets required for the above protocols are generated from calibrated beampslitter pulses with tanh-shaped ramps, where different $U_{\text{BS}}(\varphi)$ are obtained by changing the relative phase of our drives.
    \textbf{e}, The benchmarking sequences consist of randomly generated pulses that, under ideal operation, map $\vert 0_a 1_b \rangle$ back to itself.
    After each sequence, we measure whether the coupler is in its ground state, the presence of a photon in Bob, and the presence of a photon in Alice using an additional swap  gate.
    This lets us generate the raw, coupler-selected and leakage-detected RB datasets.
    All sequences are also conditioned on Bob's ancilla ending in its ground state, to discount first-order effects of ancilla heating.
}\label[fig]{fig3}
\end{figure*} 
 
\subsection*{Benchmarking fidelity in the single-photon subspace}
We precisely characterize the fidelity and noise bias of our beamsplitter by using it to implement universal control of the dual-rail qubit subspace, allowing techniques akin to standard randomized benchmarking (RB) protocols \cite{RBKnill,EmersonRBPrecursor,RBKristine}.
First, we identify the amplitude and frequency required for a fixed-length pulse (\cref{fig3}d) to achieve the beamsplitter unitary
\begin{linenomath*}
    \begin{equation}
        U_{\text{BS}}\left(\varphi\mathord{=}0\right)= e^{i\pi/4 (\hat{a}^\dagger \hat{b} + \hat{a} \hat{b}^\dagger)}
        \label[eq]{eq:Ubs},
    \end{equation}
\end{linenomath*}
by repeating the pulse to perform swaps between Alice and Bob, and iteratively checking its performance up to $\sim1000$ such repetitions (Supplementary Note 8).
This calibrates an effective $X_{\pi/2}$ gate for the dual-rail qubit, with a relative amplitude precision of less than $3\times10^{-6}$ and frequency precision of less than $3\times10^{-7}$.
Through phase control and repetition, we use this pulse to construct a set of native gates,
\begin{linenomath*}
    \begin{equation}
        G_{\text{DR}} = \{X_{\pi/2}, Y_{\pi/2}, X_{-\pi/2}, Y_{-\pi/2}, X_{\pi}, Y_{\pi}\},
        \label[eq]{eq:gate_set}
    \end{equation}
\end{linenomath*}
that generate the Clifford group for the dual rail qubit (\cref{fig3}d, Supplementary Note 9).
This allows a form of direct randomized benchmarking \cite{DirectRB}, which under uniform sampling should convert both dephasing and coherent control errors into an effective depolarization channel.
The dominant but detectable error of cavity photon loss appears as a leakage to the orthogonal state $\vert 0_a 0_b \rangle$, which is not converted to depolarization under this protocol, but can be separately quantified and selected out in post-processing.

The RB protocol consists of initializing the system in $\vert 0_a 1_b \rangle$ with the coupler and ancilla prepared in their ground states, and running sequences of varying lengths of randomly chosen gates from $G_{\text{DR}}$.
Each sequence ends with an additional gate from $G_{\text{DR}}$ that maps the state back to $\vert 0_a 1_b \rangle$, after which the presence of an excitation in the coupler, Bob, and Alice are measured (\cref{fig3}e).
We explicitly discount the effects of the ancilla in all results shown by separately measuring the ancilla and only including sequences where it ends in the ground state.
 The sequences range up to 8100 gates, where we choose up to 900 random gates (limited by FPGA memory), and repeat each gate nine times to fully capture the fidelity decay timescale (this is roughly equivalent to performing nine times as many random gates and provides a lower-bound for the fidelity of a non-repeated sequence, see Supplementary Note 9).
 We average over $\sim 10^5$ such semi-random sequences at each sequence length.
 
We observe that the average success probability of returning to Bob (\cref{fig3}b) decays exponentially, with an effective time constant of $\tau_\text{RB} = 1271\pm4$ gates.
This curve is normalized to account for effects from state preparation and readout infidelity.
We compare this curve to post-selected sequences where the coupler explicitly ended in the ground state, finding remarkable similarity.
This comparison strongly suggests that the coupler-heating induced errors on actual beamsplitter-based gates have been suppressed to below the measurement accuracy, agreeing with our earlier measurement of diminished driven coupler excitation.

Finally, we quantify our beamsplitter infidelity and noise bias by comparing the raw protocol to sequences where the system was observed to have not experienced a photon loss error. We focus on the span of the first $2250$ gates for this comparison (\cref{fig3}c, plotted on a log scale), where the raw decay of Bob's population suggests an un-selected single gate infidelity of $0.078 \pm 0.001\%$. On detecting out leakage events, our evolution changes to a curve that decays toward a steady-state fidelity of $\sim0.5$ instead of 0, as expected (Supplementary Figure 7).
For a fair comparison between this curve and the raw RB, we re-scale this fidelity to also have a steady-state of zero (Supplementary Note 10).
This helps illustrate the clear improvement under error-detection to an infidelity per gate of $0.020 \pm 0.001\%$, which shows that discarding only one out of every $\sim1300$ shots per gate can lead to a $3.9 \pm 0.2 \times$ increase in gate performance, enabled by the cavity noise-bias. We reiterate that the fact that the dominant errors remain photon loss, which is detectable or correctable with various bosonic encoding schemes, satisfies one of the key goals for a high-performance inter-cavity interaction.

Notably, because our gate-set is crafted from nearly identical beamsplitter pulses, with a random gate from $G_{DR}$ containing $4/3$ beamsplitters on average, we can directly convert these gate infidelities into an effective single beamsplitter fidelity.
 Our measurements thus imply an effective un-selected beamsplitter fidelity of $99.941 \pm 0.001\%$, which improves on leakage detection to  $99.985 \pm 0.001\%$.
The remaining errors after leakage detection can be due to intrinsic dephasing of the cavities, drifts in our control electronics, or other effects that are not treated by the  post-selection protocols, like cascaded heating and decay events of the ancilla \cite{JamesDualRailPaper}.

\section*{Discussion}

In conclusion, we constructed a microwave implementation of a tunable cavity-cavity beamsplitter and characterized its performance within the cavities' joint single-photon subspace. We obtained a beamsplitter gate fidelity exceeding $99.94\%$ in this subspace, and were limited by detectable single-photon loss in the cavities.
This performance was enabled by carefully engineering the drive frequencies and leveraging fundamental symmetries of the nonlinear converter to keep the coupling mode in its ground state even when driving a fast beamsplitter.
Such high-fidelity control of a strongly driven nonlinear element in the presence of decoherence is a significant step forward for fast parametric operations in circuit-QED.

While this system was optimized to engineer a beamsplitter between 3D cavities, the generality and versatility of the design framework  far exceed this specific application.
One could apply such a beamsplitter to on-chip resonators, or to phononic modes in hybrid architectures.
 Other low-order mixing processes allowed by the converter's symmetry, such as two-mode or single-mode squeezing, could also be implemented while still reaping the benefits of parity protection and suppressed coupler-induced infidelity.
 Parity protection is also not exclusive to the SQUID Hamiltonian \cite{JPC_Paper,ATS_Paper}, and future converter designs could leverage more advantageous forms of this selection rule.

Beyond the context of parametric interactions, we have also demonstrated the delivery of AC flux in a high-Q 3D environment.
This can be used to control other devices in similar architectures that require fast-flux modulation.
The electromagnetic simulation techniques we have developed are also readily applicable to other work involving driven circuit-QED systems, for both 3D and planar devices.

Finally, the demonstration of high-fidelity control in the dual-rail subspace motivates the hypothesis that this subspace could itself be used as a computational qubit \cite{DivincenzoDualRail,AumentadoDualRail,JamesDualRailPaper}.
The error-hierarchy of detectable decay over dephasing makes the dual-rail qubit amenable to erasure conversion \cite{ShrutiErasureConversion}, an approach potentially yielding high thresholds in the surface-code architecture.
The single-qubit control demonstrated here can be extended to realize a high-fidelity gate-set \cite{TakaPaper} for multi-qubit control, charting a path towards a general dual-rail qubit-based architecture in circuit-QED.
The performance in higher-photon manifolds is the subject of further research, with promising avenues including arraying the SQUID element, and creating parity-protected couplers with suppressed Kerr nonlinearity.

\section*{Methods}
\subsection*{Deriving the programmable beamsplitter Hamiltonian}

We now show how differential driving at the correct resonance condition generates a beamsplitter between Alice and Bob. For the sake of simplicity in derivations, we set $\hbar = 1$.
The differentially-driven SQUID Hamiltonian in \cref{eq:diff} can be expanded in the ladder operators of the bare SQUID coupler ($\hat{\tilde{c}}, \hat{\tilde{c}}^\dagger$), in a frame rotating at the coupler frequency ($\omega_c$), and gives (up to quartic order):
\begin{linenomath*}
    \begin{align}
        \hat{\mathcal{H}}_{tot}^{(4)} = &-\frac{E_C}{2} \;{\hat{\tilde{c}}^\dagger}^2 {\hat{\tilde{c}}}^2 \nonumber \\
        &+ E_J \left(\cos\left(\phi_d\right)- 1\right) \; \theta_{c,\text{zpf}}^2 \; {\hat{\tilde{c}}^\dagger} {\hat{\tilde{c}}}.
         \label[eq]{eq:freq_mod}
    \end{align}
\end{linenomath*}
Here, $\phi_d$ is assumed to be driven at frequencies far off-resonant from the coupler squeezing condition, and $\theta_{c,\text{zpf}} = \left(\frac{2E_C}{E_J}\right)^\frac{1}{4}$ is the zero-point fluctuations of the bare coupler phase.

The drive ($\phi_d$) modulates the bare coupler frequency, which enacts a beamsplitting process between the cavities Alice and Bob ($\omega_{a,b}$) through their linear capacitive couplings to the coupler ($\omega_c$):
\begin{linenomath*}
    \begin{equation}
        \hat{\mathcal{H}}_{\text{coupling}} = \sum_{\hat{\tilde{k}} = \hat{\tilde{a}}, \hat{\tilde{b}}} g_k (\hat{\tilde{k}} + \hat{\tilde{k}}^\dagger)(\hat{\tilde{c}} + \hat{\tilde{c}}^\dagger)  \nonumber
    \end{equation}
\end{linenomath*}
Here $\hat{\tilde{k}} = \hat{\tilde{a}}, \hat{\tilde{b}}$ represent the bare modes of Alice and Bob respectively.
The coupling strengths $g_{a,b}/2\pi \approx 80$ MHz are designed to place us in the dispersive regime ($\Delta_{a, b} = \omega_{a, b} - \omega_{c} \gg g_{a,b}, \; E_C$) and therefore we can follow standard circuit-QED derivations~\cite{Krantz2019,Blais2021} to diagonalize the linear part of the Hamiltonian, yielding \cref{eq:participation} in the main text.
We then move into a frame rotating at the dressed frequencies of Alice, Bob and the coupler respectively and keep relevant terms:
\begin{linenomath*}
    \begin{align}
        \hat{\mathcal{H}}_\text{tot}^{(\text{RWA})} = &\;\hat{\mathcal{H}}_\text{static}^{(\text{RWA})} + \hat{\mathcal{H}}_\text{driven}^{(RWA)} \nonumber\\
        \hat{\mathcal{H}}_\text{static}^{(\text{RWA})} = &\;-\sum_{k=a,b} 2E_C \beta_k^2 \hat{c}^\dagger \hat{c} \hat{k}^\dagger \hat{k}
        - \frac{E_C}{2} {\hat{c}^{\dagger^2}} \hat{c}^2 \nonumber \\
        \hat{\mathcal{H}}_\text{driven}^{(RWA)} = &\;\frac{\omega_c}{2} (\cos\left(\phi_d\right)-1) \times \nonumber \nonumber\\
        & \sum_{k, k'=a,b,c}\beta_k \beta_{k'} \; \hat{k}^\dagger \hat{k'} \; e^{i\left(\omega_k - \omega_{k'}\right)t}
        \label[eq]{eq:4wm_Hamiltonian}
    \end{align}
\end{linenomath*}
where $\hat{a}, \hat{b}, \hat{c}$ are the ladder operators for the dressed Alice, Bob and coupler modes respectively, $\beta_{a, b} = \frac{g_{a, b}}{\Delta_{a, b}}$ are the participations of Alice and Bob in the coupler mode, and $\beta_c \approx 1$.

Next, we introduce a pair of drive tones whose response on the actuator is given by:
\begin{linenomath*}
    \begin{equation}
        \phi_d = \sum_{1,2} \phi_{d_{1,2}}(t)  =  \sum_{1,2} \vert \phi_{d_{1,2}} \vert \sin \left(\omega_{d_{1,2}}t + \varphi_{d_{1,2}}\right)
        \label[eq]{eq:drives}
    \end{equation}
\end{linenomath*}
which are IQ-modulated signals, with Local Oscillators (LOs) that are directly derived by down-converting Bob and Alice's LOs, respectively, using a reference signal $\omega_\Delta$ (see Supplementary Note 13). This places the difference of the two drive frequencies close to the Alice-Bob detuning by design:
\begin{linenomath*}
    \begin{align}
        &\omega_{d_{2,1}} = \omega_\Delta - \omega_{a, b} + \Delta_{d_{2,1}}^{(\text{ssb})} \\
        \implies &\Delta_d = \omega_{d_2} - \omega_{d_1} = \Delta_{ab} + \Delta_{d_2}^{(\text{ssb})} - \Delta_{d_1}^{(\text{ssb})}
        \label[eq]{eq:freq_matching}
    \end{align}
\end{linenomath*}

\begin{figure*}[!b]
\includegraphics[width=0.95\textwidth]{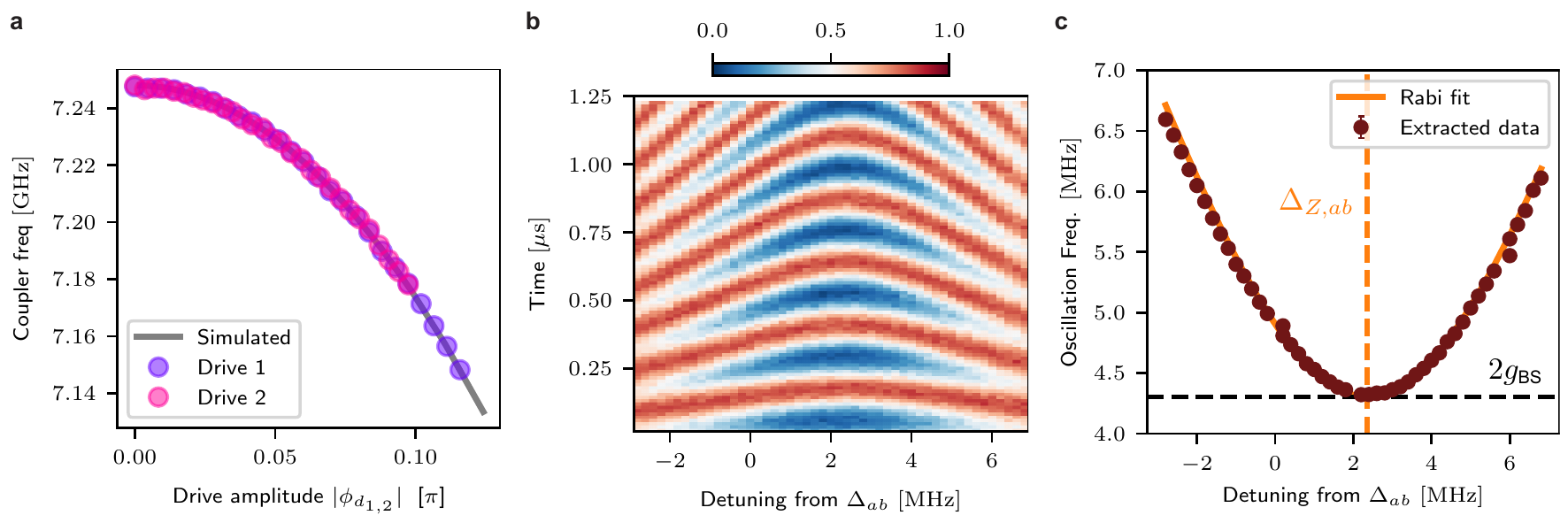}
\centering
\caption{
\textbf{Measured driven Zeeman shift and cavity swaps.}
\textbf{a}, We measure the coupler frequency through direct spectroscopy under a single drive tone, as a function drive amplitude. Comparing the Zeeman shift for either drive tone (pink, purple) to the prediction from Floquet simulation (grey solid line) allows accurate calibration of the drive strength in terms of the driven junction phase $\phi_{d1,2}$.
\textbf{b}, Measured population in Bob in the presence of both drive tones as a function of the drive-detuning ($\Delta_{d} - \Delta_{ab}$) and the time of evolution.
A single photon is prepared in Bob, and the drives swap this photon between Alice and Bob under the detuned beamsplitter interaction.
\textbf{c,} Fitting the oscillations as a function of drive-detuning to the Rabi model allows us to calibrate amplitude-dependent shift in the resonance condition ($\Delta_{Z,ab}$) and strength $g_{\text{BS}}$ of the beamsplitting interaction.}
\label[fig]{fig:Chevron_Zeeman}
\end{figure*}

Here $\omega_\Delta \approx \omega_a + \omega_b - \omega_{\text{buffer}}$ is chosen to place the the drive tones symmetrically about the buffer mode resonance, and $\Delta_{d_{1,2}}^{(\text{ssb})}$ are the detunings in the sideband modulation frequency for each drive tone with respect to the sidebands used to control Alice and Bob respectively.
We set $\Delta_{d_{2}}^{(\text{ssb})} = 0$ and use $\Delta_{d_{1}}^{(\text{ssb})} \ll \Delta_{d}$ as a finely tunable parameter to find the true beamsplitter resonance condition. Similarly, the drive phases ($\varphi_{d_{1,2}}$) are defined with respect to the control of Alice and Bob respectively, with $\varphi_{d_{2}}=0$ and $\varphi_{d_{1}} = \left\{ 0, \pi/2, \pi,-\pi/2 \right\}$ being used to generate the various pulses required for the randomized benchmarking sequence.

In this limit of $\Delta_d \approx \Delta_{ab} \ll \omega_{d_1} + \omega_{d_2}$, we can simplify the drive using the Jacobi-Anger expansion and only keep terms that are either static or rotating at the detuning frequency (up to second order in drive strength):
\begin{linenomath*}
    \begin{align}
        \cos(\phi_d) &\approx J_0\left(\vert \phi_{d_1} \vert\right) J_0\left(\vert \phi_{d_2} \vert\right)  
        \nonumber \\
        &+ 2 J_1\left(\vert \phi_{d_1} \vert\right) J_1\left(\vert \phi_{d_2} \vert\right) \cos\left(\Delta_d t + \varphi_{d_1}\right),
        \label[eq]{eq:JacobiAnger}
    \end{align}
\end{linenomath*}
where $J_0$ and $J_1$ are the zeroth and first order Bessel functions. Substituting this into \cref{eq:4wm_Hamiltonian}, we find, keeping only slow-rotating terms:
\begin{linenomath*}
    \begin{align}
        \hat{\mathcal{H}}_\text{driven}^{(\text{RWA})} \approx &\frac{\omega_c}{2} \left[J_0\left(\vert \phi_{d_1} \vert\right) J_0\left(\vert \phi_{d_2} \vert\right)-1\right] \sum_{k=a,b,c} \beta_k^2 \hat{k}^\dagger \hat{k} \nonumber \\
        &+  \frac{\omega_c}{2}J_1\left(\vert \phi_{d_1} \vert\right) J_1\left(\vert \phi_{d_2} \vert\right) \beta_a \beta_b \nonumber \\
        &\times \left(e^{i(\Delta_d - \Delta_{ab}) t + \varphi_{d1}} \hat{a} \hat{b}^\dagger + \text{h.c.}\right)
        \label[eq]{eq:H_full_driven}
    \end{align}
\end{linenomath*}
Importantly, this approximation only holds when the drive tones don't activate any other higher-order resonances. The first part of this driven Hamiltonian is a driven frequency shift, which we call the AC-Zeeman shift, and is reminiscent of the Stark shift in a regular charge-driven transmon. This is easier to see when rewriting it for small amplitude, where $\left(1-J_0(x)\right) \approx x^2/4$:
\begin{linenomath*}
    \begin{equation}
        \Delta_{\text{Z}, k} = -\beta_k^2 E_C \left( \frac{\vert\phi_{d_1}\vert^2 + \vert\phi_{d_2}\vert^2}{\theta_{c,\text{zpf}}^2 }\right)
        \label[eq]{eq:Zeeman_shift}
    \end{equation}
\end{linenomath*}
Specifically, the frequency shift of the coupler ($\Delta_{\text{Z}, c}$) provides a simple way to experimentally calibrate the strength of the drives (see \cref{fig:Chevron_Zeeman}a).

Finally, we derive the beamsplitting Hamiltonian by assuming the coupler to be in the ground state ($\langle c^\dagger c \rangle  = 0$), which eliminates contributions from $\hat{\mathcal{H}}_\text{static}^{(\text{RWA})}$.
We move into frames rotating at Zeeman-shifted frequencies for both Alice and Bob ($\omega_{ a,b}+\Delta_{Z,(a,b)}$).
We carry out an additional frame transformation for Alice, to a frame rotating at $\Delta_\text{BS} = \Delta_{ab} - \Delta_d + \Delta_{\text{Z}, ab}$, which gives us the beamsplitter Hamiltonian:
\begin{linenomath*}
    \begin{equation}
        \hat{\mathcal{H}}_\text{BS}/\hbar = \Delta_\text{BS} \hat{a}^\dagger \hat{a} + g_\text{BS}\left(e^{i\varphi_\text{BS}}\hat{a}^\dagger \hat{b} + \text{h.c.} \right).
        \label[eq]{eq:methods_bs}
    \end{equation}
\end{linenomath*}
Here,  $\varphi_\text{BS} = \varphi_{d_1}$, $\Delta_{Z,ab}=\Delta_{Z,a}-\Delta_{Z,b}$ is the relative Zeeman shift of the cavities, and $g_\text{BS}$ is the value described in the main text in \cref{eq:g_bs_value}.

We find $\Delta_{Z,ab}$ to be on the order of $g_\text{BS}$, which needs to be taken into account when finding the resonance condition. We approximately  calibrate for it by preparing a single photon in Bob and examining the system's resonant evolution as a function of $\omega_{d_1}$. This produces a Chevron-like pattern (\cref{fig:Chevron_Zeeman}b), whose oscillations follow a detuned Rabi model: $\omega_\text{osc} = \sqrt{4g_\text{BS}^2 + \Delta_\text{BS}^2}$. Fitting to this model (\cref{fig:Chevron_Zeeman}c) lets us find both the $g_\text{BS}$ at any amplitude and the drive detuning required for $\Delta_\text{BS}=0$.

\begin{figure*}[!ht]
  \centering
  \includegraphics[width=0.95\textwidth]{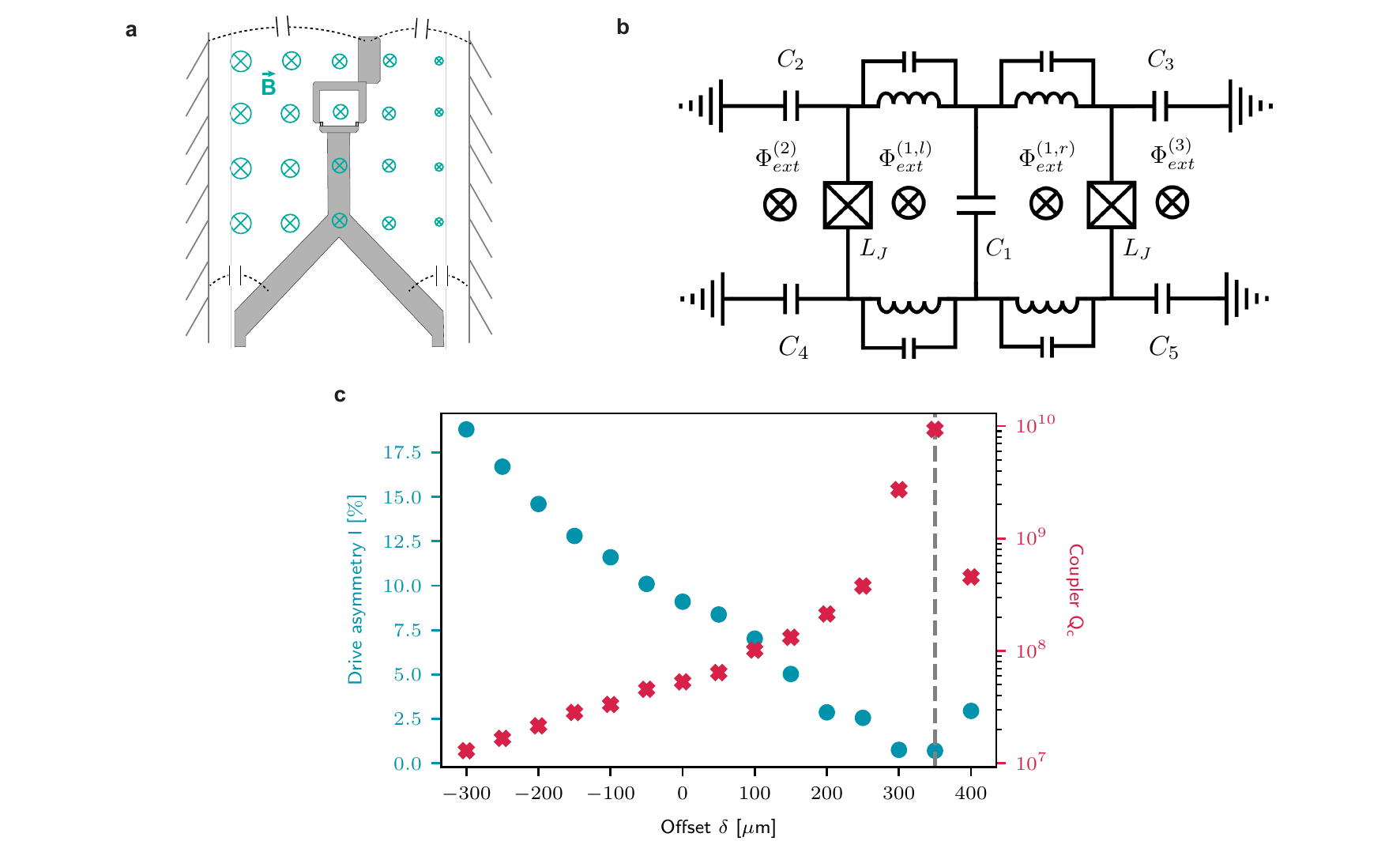}
  \caption{
  \textbf{Minimizing residual drive asymmetry in the SQUID via fine-tuning the circuit geometry.}
\textbf{a}, The SQUID device in the buffer cavity package driven by an oscillating B-field. The cylindrical geometry
of the buffer cavity (outer wall radius greater than the inner wall) dictates that the B-field has a non-uniform distribution along the radial direction (from left to right). The dashed lines with capacitor symbol represents the capacitance between the antenna pads of the SQUID and the wall of the package. 
\textbf{b}, The lumped-element circuit model of the SQUID device, taking into account the geometry of SQUID and the spatial distribution of the B field. 
\textbf{c}, HFSS simulation of the quality factor of the common mode (red) and drive asymmetry (blue, defined in \cref{eq:drive_asymmetry_HFSS}) as functions of the top pad displacement from center to right, $\delta$. As a result of introducing this asymmetry in the device geometry, the optimization of the drive asymmetry and the quality factor are simultaneously achieved at $\delta\approx350\mu$m.}
\label[fig]{HFSS_asymmetry}
\end{figure*}

\subsection*{Fine-tuning a differential drive}
We realize a purely differential drive by utilizing the separation of the coupler and actuator in electromagnetic space --- the coupler has a large electric dipole moment, while the actuator appears primarily as a magnetic dipole. This provides a convenient way to address these two modes in a 3D architecture, which is not immediately present in other parity-protected devices like the asymmetrically threaded SQUID (ATS) \cite{ATS_Paper} or the Josephson parametric converter (JPC) \cite{JPC_Paper}.
We require our drive to imitate a spatially uniform AC flux in the SQUID loop, with minimal stray capacitive coupling to its electric dipole.
The introduction of such a high-frequency flux-bias line into a superconducting package is challenging with existing techniques, since it must both preserve the lifetime of the cavities and also effectively screen parasitic driven electric fields.

We are able to conveniently address both of these issues by utilizing the natural geometric separation of the electric and magnetic fields in a $\lambda/4$ co-axial post-cavity.
We engineer our drive antenna as a 3D cavity that we call the 'buffer mode', that is incorporated into the same monolithic package that contains our high-Q storage cavities (\cref{fig1}b), thus introducing minimal additional seam or dielectric loss.
The base of this buffer cavity forms a virtual electric ground where we simultaneously have an E-field node and a B-field anti-node.
Placing the SQUID loop in this region, with its electric dipole moment oriented perpendicular to the direction of any residual E-field, automatically smothers  any stray capacitive coupling between the buffer mode and the coupler.

Any additional effects due to a spatial non-uniformity of the magnetic field threading the loop can be addressed through a purposely-engineered asymmetric capacitance in the on-chip SQUID device (\cref{HFSS_asymmetry}a).
This asymmetry in the capacitance serves as a control knob for tuning the coupling between the electromotive force and the coupler mode in the presence of an alternating magnetic field, allowing to compensate for residual drive asymmetry \cite{KochACFluxQuantization}.
To illustrate this effect, we analyze the circuit model in \cref{HFSS_asymmetry}b that captures this effect to represent the actual experimental device.
We take into account a non-uniform AC magnetic field that is distributed across the device, not only inducing a flux in the SQUID loop, but also inducing an electromotive force on the shunting capacitors.
Assuming the geometric inductance of the loop is much smaller than the Josephson inductance (Supplementary Note 1), we arrive at the effective Hamiltonian of
\begin{linenomath*}
    \begin{equation}
        \hat{\mathcal{H}}=\frac{\left(\hat{Q}_c-C_\Sigma V_{\text{emf}}\right)^2}{2C_\Sigma}-E_{J}\cos{\frac{\Phi_\text{ext}^{\left(1,\Sigma\right)}}{2\phi_0}}\cos{\frac{\hat{\Phi}_c}{\phi_0}}.
    \label[eq]{eq:SQUID_Lagrangian_4}
    \end{equation}
\end{linenomath*}
Here, $\hat{Q}_c$ is the charge operator of the common mode, $C_\Sigma=C_1+C_2+C_3+C_4+C_5$ is its total shunting capacitance, and $\Phi_\text{ext}^{\left(1,\Sigma\right)}=\Phi_\text{ext}^{\left(1,l\right)}+\Phi_\text{ext}^{\left(1,r\right)}$ is the total external flux penetrating the SQUID loop. The charge operator $\hat{Q}_c$ and the flux operator $\hat{\Phi}_c$ are related to the Cooper-pair number operator $\hat{n}_c$ and the superconducting phase operator $\hat{\theta}_c$ through $\hat{Q}_c = 2e\hat{n}_c$, $\hat{\Phi}_c = \phi_0\hat{\theta}_c$, where $e$ and $\phi_0$ are the electron charge and the reduced flux quantum, respectively.
The electromotive voltage, $V_\text{emf}$, is given by
\begin{linenomath*}
    \begin{align}
    V_\text{emf}&=\frac{C_1}{2C_\Sigma}\dot{\Phi}_\text{ext}^{\left(1,\delta\right)}+\frac{C_l}{C_\Sigma} \left(\frac{1}{2}\dot{\Phi}_\text{ext}^{\left(1,\Sigma\right)}+\dot{\Phi}_\text{ext}^{\left(2\right)}\right) \nonumber\\
    &-\frac{C_r}{C_\Sigma} \left(\frac{1}{2}\dot{\Phi}_\text{ext}^{\left(1,\Sigma\right)}+\dot{\Phi}_\text{ext}^{\left(3\right)}\right), \label[eq]{eq:V_emf}
    \end{align}
\end{linenomath*}
with $\Phi_\text{ext}^{\left(1,\delta\right)}=\Phi_\text{ext}^{\left(1,l\right)}-\Phi_\text{ext}^{\left(1,r\right)}$, $C_l = C_2+C_4$, $C_r = C_3+C_5$.

Shifting the upper antenna pad of the SQUID from the center to the right simultaneously increases $C_3$ and decreases $C_1$ and $C_2$ in Eq.~\ref{eq:V_emf}.
Fine-tuning this offset ($\delta$) can result in the minimization of $V_\text{emf}$, as demonstrated in \cref{HFSS_asymmetry}c, where we simulate the drive asymmetry $l$ (Supplementary Note 2) at various offsets in Ansys finite-element High-Frequency Simulation Software (HFSS).
We obtain this asymmetry from the currents across the two junctions,
\begin{linenomath*}
    \begin{equation}
    l=\frac{\left\vert\phi_c\right\vert}{\sqrt{\left\vert\phi_c\right\vert^2+\left\vert\phi_d\right\vert^2}}=
    \frac{\left\vert I_{J1}+I_{J2} \right\vert}{\sqrt{2\left[\left\vert I_{J1} \right\vert^2+\left\vert I_{J2} \right\vert^2\right]}},
    \label[eq]{eq:drive_asymmetry_HFSS}
    \end{equation}
\end{linenomath*}
with $I_{J1,2} =\phi_0\phi_{1,2}/L_{J1,2}$. 
When the pad is shifted towards the right by $\delta\approx 350\mu$m, the asymmetry in the drive is almost completely eliminated, resulting in the desired purely differential drive.
At the same time, the quality factor of the common mode (coupler) reaches a maximum, since the Purcell effect from the coupler's linear coupling to the buffer mode is highly suppressed.
In the experiment, this lets us strongly couple the buffer mode to a coaxial drive line and achieve a quality factor ($\mathsf{Q}$) as low as 12 while maintaining a $\mathsf{Q}_\text{c} \sim 2\times 10^6$ for the coupler, and $\mathsf{Q}_\text{a} \sim 1.4\times 10^7$, $\mathsf{Q}_\text{b}\sim 1.2\times 10^7$ for Alice and Bob respectively. We also make sure that other types of back-actions of the buffer mode, such as the coherent- or thermal-photon-induced dephasing, is not limiting the coherence of the coupler and the high-Q storage modes (Supplementary Note 3).

\subsection*{Experimentally characterizing residual drive asymmetry}
Non-idealities in the experimental implementation of our differential drive inevitably result in a small but finite common-mode drive strength.
One way to experimentally characterize this residual drive asymmetry is to probe the strength of interactions that are uniquely generated by the common-mode drive.
A candidate for this is the `ge/3' interaction, where a single drive tone at $\approx \omega_{c}/3$ drives Rabi oscillations of the coupler between its ground state and its excited state through three drive photons.
The rate of this Rabi-like process is given by (Supplementary Note 1):
\begin{linenomath*}
    \begin{equation}
    \Omega_{\text{ge/3}}=\frac{E_J}{\hbar} \;\theta_{c,\text{zpf}} \phi_c \left(\vert \phi_c\vert ^2/3 + \vert \phi_d \vert ^2\right). \label[eq]{ge_3_rate}    
    \end{equation}
\end{linenomath*}

\begin{figure}[ht]
\includegraphics[width=0.48\textwidth]{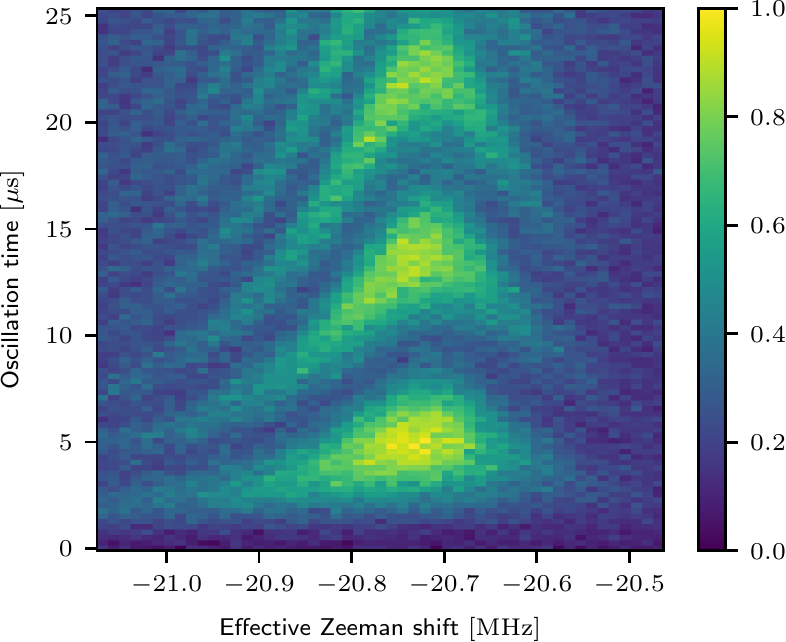}
\caption{
\textbf{Experimentally characterizing drive asymmetry.}
We directly probe the coupler `ge/3' transition due to any residual common-mode drive. The coupler is prepared in the ground state at $t=0\mu s$, followed by a single-tone drive through the buffer-mode near $\omega_c/3$. The rate of coherent oscillation of the coupler population (color-scale) compared to the effective Zeeman shift (x-axis) of this resonance frequency bounds the drive asymmetry to $<1\%$ }
\label[fig]{fig:Meth_ge3}
\end{figure}

To obtain the relative drive asymmetry, we compare the strength of the ge/3 process to the Zeeman-shift induced by the same single-tone drive, which depends on both $\phi_c$ and $\phi_d$ :\
\begin{linenomath*}
    \begin{align}
        \Delta_{\text{Z},c}&\approx \left[J_0\left(\vert\phi_c\vert\right)J_0\left(\vert\phi_d\vert\right)-1\right]\omega_c/2\nonumber\\
        &\approx -\frac{\omega_c}{8}\left(\lvert\phi_c\rvert^2+\lvert\phi_d\rvert^2\right).
        \label[eq]{ge_3_Zeeman}
    \end{align}
\end{linenomath*}
Note the difference between this and \cref{eq:Zeeman_shift} which assumes a purely differential drive with two tones.
Measuring the strength of both these processes (\cref{ge_3_rate,ge_3_Zeeman}) is sufficient to extract $\vert \phi_c \vert$ and $\vert\phi_d\vert$.
As illustrated in \cref{fig:Meth_ge3}, we observe the $ge$/3 Rabi oscillation of the coupler with a rate of $\Omega_{\text{ge/3}}/2\pi = 0.12$ MHz.
The resonant frequency of the $ge$/3 oscillation is at $2.408$ GHz, which  corresponds to an effective Zeeman shift of $\Delta_{\text{Z},c}/2\pi = -21$MHz. 
Using Eq.~\ref{ge_3_rate} and~\ref{ge_3_Zeeman}, we obtain $\lvert\frac{\phi_c}{\phi_d}\rvert \approx 2.5e-3$, corresponding to a drive asymmetry of $l < 1\%$.
It is possible that the drive asymmetry is a little higher in the actual experiment, due to a deviation from our assumption of perfectly symmetric junction energies, but we expect this deviation to be small (Supplementary Note 7).

\subsection*{Decoherence in the dual-rail subspace}   
We can derive the expected evolution and the beamsplitter fidelity from the time evolution of the density matrix in the dual-rail subspace (Supplementary Note 4):
\begin{linenomath*}
    \begin{align}
    \rho_{\text{DR}}(t) &=\frac{1}{2} e^{-\kappa_1 t} \; \times\nonumber\\
    &
    \begin{bmatrix}
    1+e^{-\kappa_{\varphi}t}\cos\left(\Omega t\right) & i\sin\left(\Omega t\right) e^{-\kappa_\varphi t}\\ 
    -i\sin\left(\Omega t\right) e^{-\kappa_\varphi t} & 1-e^{-\kappa_{\varphi}t}\cos\left(\Omega t\right)
    \end{bmatrix},
    \label[eq]{eq:dm_dual_rail}
    \end{align}
\end{linenomath*}
where we have used $\Omega = 2 g_{\text{BS}}$ for simplicity.
The upper left element measures the probability of finding a photon in Bob, which is Eq. \ref{eq:Pbob} in the main text.

To calculate the beamsplitter fidelity, we can set the evolution time to be $t_\text{BS} = \frac{\pi}{4g_\text{BS}}$ and calculate the overlap with a lossless evolution ($\rho_{\vert\psi^{\pm}\rangle}^{(0)}$) that has no leakage or dephasing:
\begin{linenomath*}
    \begin{align}
        \mathcal{F}(t_\text{BS}) &= \text{Tr}\left[\rho_{\text{DR}}(t_\text{BS}) \times \rho_{\text{DR}}^{(0)} (t_\text{BS}) \right] \nonumber \\
        &= \frac{1}{2}\left[e^{-\kappa_1 t} + e^{-\left(\kappa_1 + \kappa_\varphi \right)t}\right] \nonumber \\
        &\approx 1 - \kappa_\text{BS} t_\text{BS}.
    \label[eq]{eq:bm_fid}
    \end{align}
\end{linenomath*}
Here $\kappa_\text{BS} = \kappa_1 + \frac{\kappa_\varphi}{2}$, which gives us Eq. \ref{eq:fidbs} in the main text.

\begin{figure}[ht]
\includegraphics[width=0.48\textwidth]{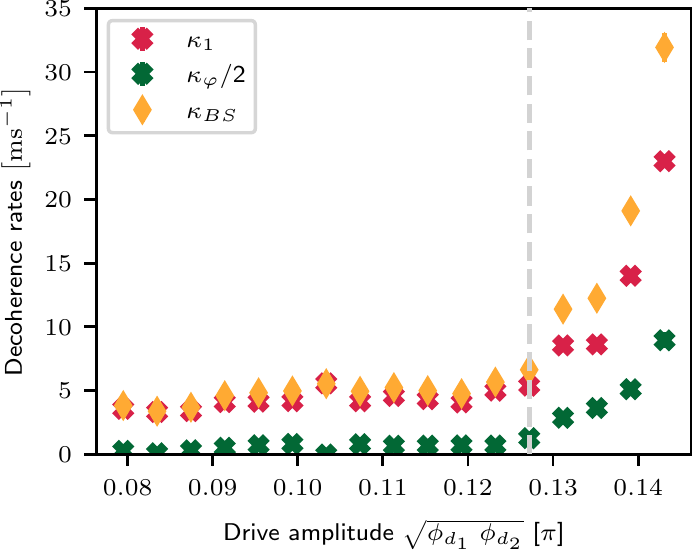}
\caption{
\textbf{Measured driven decoherence.}
Decoherence rates are extracted from short sections of the long-time evolution under both drive tones, as a function of drive amplitude. The sideband collision of coupler and Bob (see Supplementary Note 5) clearly limits the fidelity at high amplitudes, and we operate on the boundary of this collision (grey dashed line), where $\kappa_{\text{BS}}$ is limited by $\kappa_1$.
}
\label[fig]{fig:Meth_kappaBS}
\end{figure}

In order to efficiently extract $\kappa_1$ and $\kappa_{\phi}$ from the experiment, we create a single photon in Bob, and measure Bob's population under the beamsplitter interaction within 
two sections of time: $t\in[0,\:\delta t]$, and $t\in[T,T+\:\delta t]$. 
For small $\delta t$ and large $T$, the measurement result within these time windows can be approximated as
\begin{linenomath*}
    \begin{align}
        P_0(t) &= A\left(1 + \cos{(2 g_{\text{BS}} t+\phi_0)}\right) + B,\nonumber\\
        P_T(t) &= Ae^{-\kappa_1 T} \left(1 + e^{-\kappa_\varphi T} \cos{(2g_{\text{BS}} t+\phi_0)} \right) + B.
        \label[eq]{eq:short_sections}
    \end{align}
\end{linenomath*}
Here, $A$ and $A+B$ represent the amplitude and the offset of the oscillation, which are close to but not strictly equal to 0.5 in the presence of state preparation and measurement (SPAM) errors. From fitting to $P_0(t)$ we extract $A$ and $B$, which are then used in the fitting of $P_T(t)$ to find $\kappa_1$ and $\kappa_{\varphi}$. We choose $T=25\mu$s and $\delta t \approx 20 t_{\text{BS}}$ to extract the decay rates at different drive amplitudes, shown in \cref{fig:Meth_kappaBS}.

\section*{Data availability}
The data  presented in this study is available at DOI: \href{10.6084/m9.figshare.23589144.v1}{https://doi.org/10.6084/m9.figshare.23589144.v1} and more detailed source data is available from the corresponding authors upon request.

\section*{Code availability}
All computer codes used in this study are available from the corresponding authors upon request.


\bibliography{sn-bibliography}
\bibliographystyle{sn-mathphys}

\section*{Acknowledgments}
We thank J. Venkatraman and and X. Xiao for helpful conversations on parasitic nonlinear resonances. We are grateful to J. Curtis, A. Koottandavida and I. Tsioutsios for technical assistance and providing useful code. We thank B. Chapman for providing parametric amplifiers used in this experiment, and A. Read for help with DC-flux line wiring. We are grateful to C. Wehr and H. M. Moseley for help with designing substrate clamps and K. Chou for advice on beamsplitter control. We thank A. Miano for help simulating the dc-flux spectroscopy of the SQUID. We also thank P. Winkel, M. Devoret, B. Chapman, S. de Graaf, S. Xue, J. D. Teoh, T. Tsunoda, S. Chakram and J. Huang for useful discussions. This research was sponsored by the Army Research Office (ARO) under grant nos.  W911NF-16-1-0349, W911NF-18-1-0212 and W911NF-22-1-0053. The views and conclusions contained in this document are those of the authors and should not be interpreted as representing the official policies, either expressed or implied, of the Army Research Office (ARO) or the US Government. Fabrication facilities 
use was supported by the Yale Institute for Nanoscience and Quantum Engineering (YINQE) and the 
Yale SEAS Cleanroom.

\section*{Author contributions}
Y.L., A.M. designed and conducted the experiment, under the supervision of R.J.S.. Y.L., A.M. and S.G. designed the experimental device and developed the differential drive scheme. Y.L. fabricated the DC-SQUID and ancilla transmon chip devices with help from S.G. and L.F.. A.M., Y. L. and J.G. conducted measurements and analyzed the data. J.G., A.M. and Y.L. implemented the dual-rail qubit randomized benchmarking. Y.Z. and J.C. provided theoretical support for analyzing parametric processes and randomized benchmarking, respectively. S.M.G. provided important theoretical insights during the writing of the manuscript. A.M., Y.L., J.G. and R.J.S. wrote the manuscript with feedback from all co-authors.

\section*{Competing Interests}
L.F. and R.J.S. are founders and shareholders of Quantum Circuits, Inc (QCI).
S.M.G. receives consulting fees and is an equity holder in QCI.
The remaining authors declare no competing interests.

\newpage
\clearpage
\newpage

\renewcommand{\thesection}{Supplementary Note \arabic{section}} 

\renewcommand{\thefigure}{S\arabic{figure}} 
\renewcommand{\thesubsection}{\thesection.\Roman{subsection}}
\renewcommand{\thetable}{S\arabic{table}} 

\titleformat{\section}[block]
{\normalfont\bfseries}
{ \thesection \ --- }{0pt}{}

\twocolumn[{
    \centering
    \Large \textbf{Supplementary Information} \\
    \vspace{1em}
}]

\section{Hamiltonian of the driven DC SQUID}
\begin{figure}[ht]
  \centering \includegraphics[width=0.9\columnwidth]{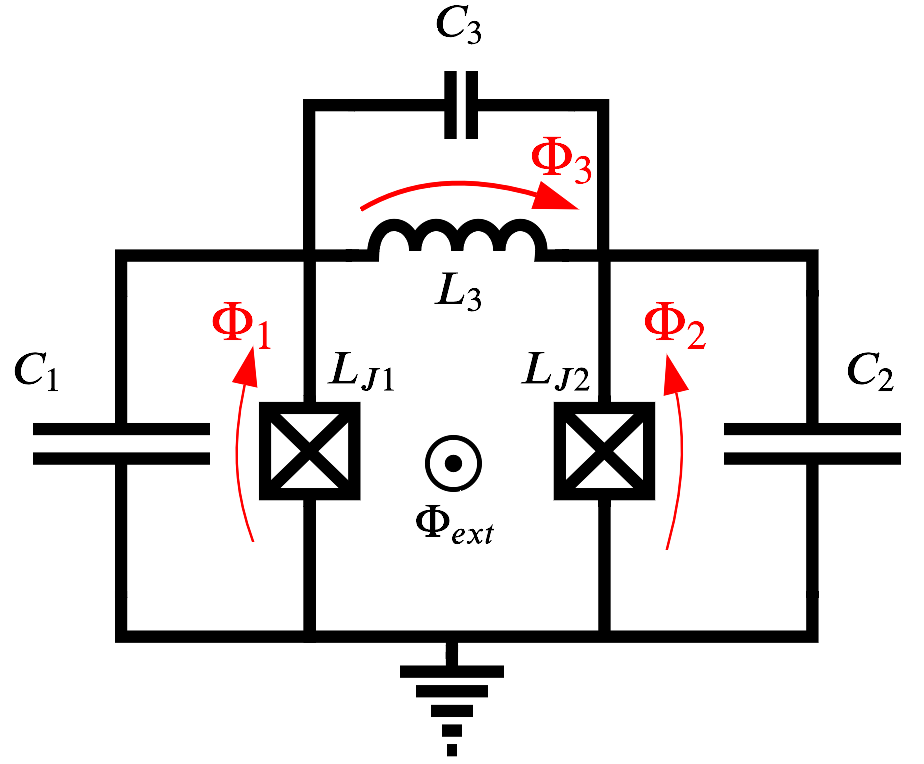}
  \caption{{\textbf{The circuit diagram of a flux-driven SQUID.}}}
\label{fig:SQUID_circuit_diagram}
\end{figure}

The two-mode nature of a DC-SQUID has been well-studied in literature \cite{SQUIDModesLecocq2011,KamalSQUIDModes}.
In this section, we derive the joint Hamiltonian for these two modes in the presence of an AC-flux modulation.
The SQUID circuit is shown in Fig.~\ref{fig:SQUID_circuit_diagram}, where we include a small but finite linear inductance $L_{3}$ and capacitance $C_{3}$ that connects the two junctions. The flux variables across the junctions and the linear inductance are labeled as $\Phi_{1,2,3}$, where the first two directly correspond to the Josephson phases $\hat{\theta}_{1,2}$ in the main text. From this general formalism, we will derive the parity-protected Hamiltonian as a special case, with $\hat{\theta}_c$ and $\phi_d$ representing the coupler and actuator respectively. We will also derive the other limit of the common-driven SQUID and explicitly establish the Hamiltonian under a continuous spectrum of drive asymmetry.

We start with the general circuit Lagrangian for the driven SQUID in Fig. \ref{fig:SQUID_circuit_diagram}
\begin{linenomath*}
    \begin{align}
    \mathscr{L}=\frac{1}{2}\sum_{i=1}^{3}C_{i}\dot{\Phi}_i^2+\sum_{i=1}^{2}E_{Ji}\cos{\frac{\Phi_i}{\phi_0}}-\frac{\Phi_3^2}{2L_{3}},\label{eq:SQUID_Lagrangian}
    \end{align}
\end{linenomath*}
with a constraint from the flux quantization relationship,
\begin{linenomath*}
    \begin{equation}
        \Phi_1-\Phi_2+\Phi_3=\Phi_{\text{ext}}+n\Phi_0,
    \end{equation} 
\end{linenomath*}
where $\Phi_0\; (\phi_0)$ is the (reduced) flux quantum.
Therefore, we may rewrite Eq.~\ref{eq:SQUID_Lagrangian} as
\begin{linenomath*}
\begin{align}
\mathscr{L}&=\frac{1}{2}C_{1}\left(\dot{\Phi}_c+\dot{\Phi}_d\right)^2+\frac{1}{2}C_{2}\left(\dot{\Phi}_c-\dot{\Phi}_d\right)^2 \nonumber\\
&+E_{J_1}\cos{\frac{\Phi_c+\Phi_d}{\phi_0}}+E_{J_2}\cos{\frac{\Phi_c-\Phi_d}{\phi_0}}\nonumber\\
&+\frac{1}{2}C_{3}\left(2\dot{\Phi}_d-\dot{\Phi}_{\text{ext}}\right)^2-\frac{\left(2\Phi_d-\Phi_{\text{ext}}\right)^2}{2L_{3}}
.\label{eq:SQUID_Lagrangian_2}
\end{align}
\end{linenomath*}
Here, we have defined the ``differential mode'' $\Phi_d$, and the ``common mode'' $\Phi_c$, as
\begin{linenomath*}
    \begin{equation}
    \Phi_{c,d}=\frac{1}{2}\left(\Phi_1\pm\Phi_2\right). \label{eq:Phi_C_Phi_D}
    \end{equation}
\end{linenomath*}
Under the assumption that the linear inductance $L_{J3}$ and the capacitance $C_{J3}$ are small, the differential mode frequency is much higher than that of the common mode. This means that, even under circuit asymmetry such that $C_{1}\neq C_{2}$ and $E_{J_1}\neq E_{J_2}$, $\Phi_c$ and $\Phi_d$ remain good approximations for the normal-mode flux variables of the circuit. From this, the equation of motion for $\Phi_d$ can be approximated as
\begin{linenomath*}
\begin{align}
&\frac{d}{dt}\frac{\partial \mathscr{L}}{\partial \dot{\Phi}_d}-\frac{\partial \mathscr{L}}{\partial \Phi_d}\approx \nonumber\\
&C_{d\Sigma}\ddot{\Phi}_d+L_{d\Sigma}^{-1}\Phi_d-2C_{3}\ddot{\Phi}_{ext}-2\frac{\Phi_{\text{ext}}}{L_{3}}=0, \label{eq:Phi_D_EOM}
\end{align}
\end{linenomath*}
where $C_{d\Sigma}=C_{1}+C_{2}+4C_{3}$ and $L_{d\Sigma}=\left(L_{J1}^{-1}+L_{J2}^{-1}+4L_{3}^{-1}\right)^{-1}$ are the total capacitance and inductance of the differential mode. When the drive frequency is much lower than the differential mode frequency, the steady state solution of Eq.~\ref{eq:Phi_D_EOM} is given by
\begin{linenomath*}
    \begin{equation}
    \left<\Phi_{d}\right>\approx\Phi_{\text{ext}}/2. \label{eq:Phi_D_expectation}
    \end{equation}
\end{linenomath*}
This can also be seen from the fact that in Eq.~\ref{eq:SQUID_Lagrangian_2}, the last term is a high energy term that requires its numerator to be a ``frozen'' degree of freedom, $\left<2\dot{\Phi}_d-\dot{\Phi}_{ext}\right>$=0. Plugging this back in to Eq.~\ref{eq:SQUID_Lagrangian_2} yields
\begin{linenomath*}
\begin{align}
\mathscr{L}&=\frac{1}{2}C_{1}\left(\dot{\Phi}_c+\frac{\dot{\Phi}_{ext}}{2}\right)^2+\frac{1}{2}C_{2}\left(\dot{\Phi}_c-\frac{\dot{\Phi}_{ext}}{2}\right)^2\nonumber\\
&+E_{J_1}\cos{\frac{\Phi_c+\frac{\Phi_{\text{ext}}}{2}}{\phi_0}}+E_{J_2}\cos{\frac{\Phi_c-\frac{\Phi_{\text{ext}}}{2}}{\phi_0}}. \label{eq:SQUID_Lagrangian_3}
\end{align}
\end{linenomath*}

In reality, the differential flux modulation may also be contaminated by stray voltage coupling of the common mode to the flux line, as well as by the non-uniformity in the spatial distribution of the time-dependent flux that creates uneven voltage drops (electromotive forces) across the shunting capacitors (see Methods ``Fine-tuning a differential drive" for a detailed analysis). Without loss of generality, the Hamiltonian can be written as 
\begin{linenomath*}
\begin{align}
\hat{\mathcal{H}}&=4 E_C \hat{n}_c^2+\epsilon\left(t\right)\hat{n}_c\nonumber\\
-&E_{J_1}\cos{\left(\hat{\theta}_c+\frac{\Phi_{\text{ext}}}{2\phi_0}\right)}-E_{J_2}\cos{\left(\hat{\theta}_c-\frac{\Phi_{\text{ext}}}{2\phi_0}\right)} , \label[eq]{eq:SQUID_Hamiltonian_general}
\end{align}
\end{linenomath*}
where $\epsilon\left(t\right)$ represents the effective common mode drive strength, and $\hat{n}_c$, $\hat{\theta}_c$ are the Cooper-pair number operator and the superconducting phase operator of the common mode, respectively. In the displaced frame, this Hamiltonian becomes (for symmetric SQUID of $E_{J_1}=E_{J_2}=E_J/2$)
\begin{linenomath*}
\begin{align}
\hat{\mathcal{H}}_{\text{disp}}&=4 E_C \hat{n}_c^2\nonumber + E_{J}\frac{\hat{\theta}_c^2}{2} \\
&-E_{J}\cos_{\text{NL}}{\left(\hat{\theta}_c+\phi_1\right)}-E_{J}\cos_{\text{NL}}{\left(\hat{\theta}_c+\phi_2\right)}\nonumber\\
&=4 E_C \hat{n}_c^2 - E_{J} \cos\phi_c \cos\phi_d\cos\hat{\theta}_c \nonumber\\
&+ E_J\sin\phi_c\cos\phi_d\sin\hat{\theta}_c - E_J \phi_c\hat{\theta}_c,
\label[eq]{eq:SQUID_Hamiltonian_general_displaced}
\end{align}
\end{linenomath*}
where $\cos_\textup{NL}x=\cos{x}+\frac{x^2}{2}$, and $\phi_{c,d}=\frac{1}{2}\left(\phi_1\pm\phi_2\right)$ are the displacements in the phases of the common and the differential mode. From \cref{eq:SQUID_Hamiltonian_general} and \cref{eq:SQUID_Hamiltonian_general_displaced}, we obtain these displacements as $\phi_d=\Phi_{\text{ext}}/2\phi_0$, and $\phi_c=\epsilon(t)\omega_d\hbar^{-1}/\left(\omega_d^2-\omega_c^2\right)$, where $\omega_c$ and $\omega_d$ are the frequencies of the coupler and the drive.  
A pure differential drive corresponds to $\phi_c=0$, where Eq.~\ref{eq:SQUID_Hamiltonian_general_displaced} reduces to
\begin{linenomath*}
    \begin{equation}
    \hat{\mathcal{H}}_{DDS} =4 E_C \hat{n}_c^2-E_{J}\cos\phi_d\cos{\hat{\theta}_c},\label{eq:SQUID_Hamiltonian_parity_protected}
    \end{equation}
\end{linenomath*}
which is the parity-protected Hamiltonian (Eq. 1) in the main text. In contrast, $\phi_d = 0$ happens for a purely charge-driven transmon, 
\begin{linenomath*}
    \begin{align}
    \hat{\mathcal{H}}_{trans}&=4 E_C \hat{n}_c^2-E_J\cos\phi_c\cos\hat{\theta_c}\nonumber\\
    &+E_J\sin\phi_c\sin\hat{\theta}_c-E_J\phi_c\hat{\theta}_c.\label{eq:SQUID_Hamiltonian_charge_driven_displaced_2}
    \end{align}
\end{linenomath*}

With Eq.~\ref{eq:SQUID_Hamiltonian_general_displaced}, we can express the strengths of some of the parametric processes in terms of the displacement amplitudes, $\phi_c$ and $\phi_d$. For example, the beamsplitter interaction arises from the $\cos\hat{\theta}_c$ modulation in the third line of Eq.~\ref{eq:SQUID_Hamiltonian_general_displaced}. For  purely RF drives where $\phi_d$ contains no DC component, both the beamsplitter rate and the Zeeman shift are  proportional to $\phi_c^2+\phi_d^2$ under weak drive strengths. On the other hand, the last line of Eq.~\ref{eq:SQUID_Hamiltonian_general_displaced} gives rise to the odd parity interactions, with the leading order term being the `$ge/3$' interaction, 
\begin{linenomath*}
\begin{align}
    \hat{\mathcal{H}}_{ge/3} &\approx E_J \left(\phi_c-\frac{\phi_c^3}{6}\right) \left(1-\frac{\phi_d^2}{2}\right)\hat{\theta}_c - E_J \phi_c\hat{\theta}_c \nonumber \\
    &\approx -\frac{E_J}{2}\phi_c\left(\phi_d^2+\frac{\phi_c^2}{3}\right)\theta_{c,\text{zpf}}\left(\hat{c}^\dagger+\hat{c}\right),
\end{align}
\end{linenomath*}
which we used to experimentally characterize the residual drive asymmetry (Methods ``Experimentally characterizing residual drive asymmetry").

\section{Parity protection and drive-frequency engineering}

In this section, we discuss the importance of parity protection and drive-frequency engineering in the differentially-driven-SQUID (DDS), by contrasting it to a charge-driven transmon. In the latter's case (Eq.~\ref{eq:SQUID_Hamiltonian_charge_driven_displaced_2}), only the `even parity' part of the Hamiltonian ($\hat{\theta}_c^{2n}$, coming from the second term) that allows even numbers of mode quanta to interact, is useful for the bilinear beamsplitting process ($\propto\hat{\theta}_c^{2}$), 
The odd-parity terms ($\hat{\theta}_c^{2n+1}$, from the second line) introduce a set of nonlinear processes that are not only unhelpful, but can be actively harmful, leading to an incoherent excitation of the coupler that dephases the beamsplitter process, or an unintended exchange of sensitive quantum information with other modes in the system or the environment.

We are able to address both of these issues by utilizing the orthogonality of the coupler and actuator in the differentially-driven SQUID (DDS, Eq.~\ref{eq:SQUID_Hamiltonian_parity_protected}), which offers a two-fold advantage through its parity protection and simpler drive engineering.

First, the forbidden linear coupling between the drive and the coupler avoids its displacement that converts dephasing into heating.
This can be intuitively understood as high-frequency coupler dephasing combining with the drive to provide excitation at the coupler frequency.
This process should be sensitive only to the spectrum of coupler frequency noise at the detuning between the drive and the coupler mode, and thus driving far-detuned from the coupler is also helpful.

Second, incoherent excitations may also result from driven resonant processes in combination with coupler decay. It is therefore imperative to avoid these resonances, the strongest of which is a squeezing of the coupler that is resonant near the coupler mode frequency (specifically, at half the g-f transition frequency).
Intuitively, this can be thought of as a coherent excitation from the ground to the second excited state, combined with an incoherent decay back to the first excited state, thus leading to an over-all incoherent heating.
This is a common issue in both the charge-driven transmon as well as the DDS, since squeezing and higher-order even-parity interactions are not forbidden by Eq. \ref{eq:SQUID_Hamiltonian_parity_protected}. However, in the DDS, the orthogonality between the actuator and the coupler greatly facilitates the engineering of the drives that are far-detuned from the coupler, suppressing both the squeezing and the dressed dephasing.

Finally, the selection rule enforced by the parity-protected Hamiltonian should forbid any odd-parity exchanges for modes participating in the couple, which is especially useful since strong drives can cause frequency shifts of the resonance condition.
This includes resonances that could either cause undesired exchanges between the cavity ($\hat{a}$) and the coupler ($\hat{c}$), like $\phi_d \;\hat{a} \hat{c}^{\dagger2} + \text{h.c.}$ which was a limiting factor in \cite{Yvonne2018_PRX_ProgrammableInterference}, or resonances that cause direct excitations of the coupler through terms like $\phi_d^3 \;\hat{c}^\dagger + \text{h.c.}$.

Which of these effects dominate in the experiment depends strongly on the frequencies of interest and the spectrum of environmental noise that couples to the system. Here, we take a broader approach by using numerical simulations to illustrate the clear difference between a common and a differential drive for a DC-SQUID, in the presence of one or both of the drive tones.
We expect the full dynamics in the presence of both cavities and a colored spectrum of environmental noise to be significantly more complicated.

\begin{figure*}[!ht]
\centering
\includegraphics[width=0.98 \textwidth]{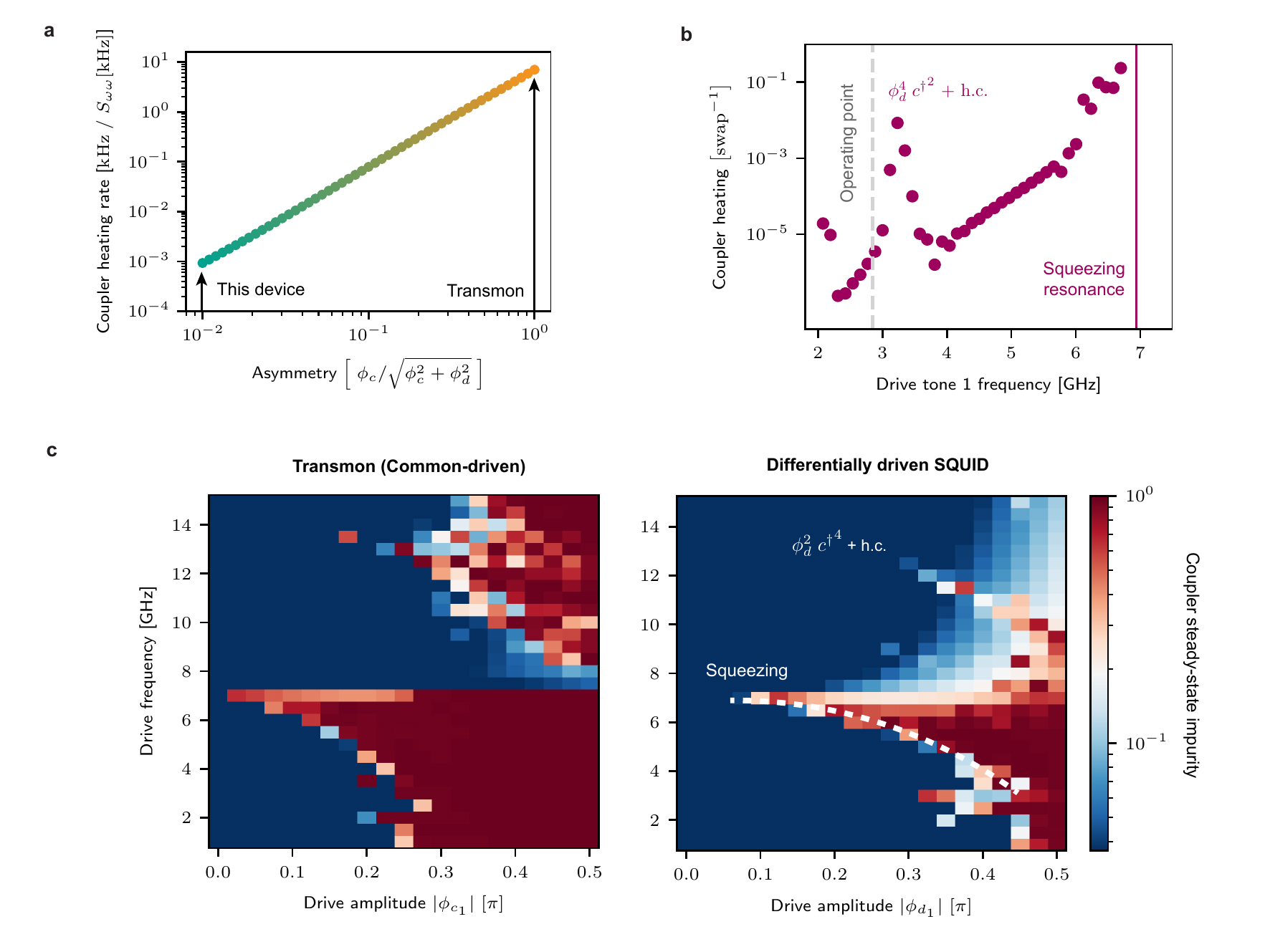}
\caption{ \textbf{Floquet simulations demonstrating advantages of the differentially-driven SQUID.}
\textbf{a}, The heating rate of the coupler from the Floquet ground state to the Floquet excited state, as a result of two drive tones ($\omega_{d_1} = 2\pi\times 2.772$GHz, $\omega_{d_2} = 2\pi\times 3.003$GHz) with various amount of drive asymmetry. Plotting the heating rate normalized by the spectral density of the dephasing $S_{\omega_c \omega_c}[\Delta_{d_2, c}]$ reveals the strong dependence of the coupler heating on the drive asymmetry, with a suppression by several orders of magnitude as we move from large asymmetries (orange) to small asymmetries (teal) that are practically achievable (the black arrows indicate the value of asymmetry in this experiment, and in the case of a single-junction transmon, respectively).
\textbf{b}, A two-tone Floquet-Markov simulation illustrating the choice of drive frequencies for the differentially driven SQUID, at a fixed drive amplitude.
The driven coupler heating rate is plotted (magenta) as a function of frequency of drive tone 1 (with the other drive frequency fixed to satisfy the beamsplitter resonance condition), in the presence of intrinsic coupler decay.
We see that even for a purely differential drive, the amplitude-dependent squeezing (here resonant at $\omega_{gf/2} = 6.94$ GHz) can convert decay into heating (magenta).
We place the drives far red-detuned from the coupler (grey dashed line), avoiding both the coupler squeezing and weaker nearby resonances, like the $\phi_d^4 \; \hat{c}^{\dagger^2}+\text{h.c.}$ process.
\textbf{c}, The impurity of the driven steady-state of the coupler, under a single drive tone, for the common and differentially driven cases.
The blue and white regions imply a cold and hot coupler respectively, while the red regions are a result of chaotic behaviour or coupler ionization.
The differentially-driven SQUID shows a significantly larger available drive space, where one can drive at higher amplitudes at almost every drive frequency without ionizing the coupler.
Some residual limiting features in this plot are also clearly recognizable, like coupler squeezing (white dashed line is a guide to the eye), and the $\phi_d^2 \hat{c}^{\dagger^4}+\text{h.c.}$ process, both of which follow a downward trend due to the Zeeman shift.
}\label[fig]{floquet_sims}
\end{figure*}

Our simulations use well-established Floquet-Markov methods \cite{HanggiFloquet,Verney2019,Yaxing_PRA_BilinearModeCoupling}.
The Floquet basis provides an appropriate basis for analyzing the driven coupler, with the Floquet ground state smoothly mapping to the undriven ground state.
Intrinsic coupler decay and dephasing can result in an effective excitation out of this ground state, represented by a `hopping' rate to other Floquet states \cite{Yaxing_PRA_BilinearModeCoupling}.
This can cause incoherent shifts in the beamsplitting strength and resonance condition, and thus dephase the beamsplitting process.
Quantifying the coupler heating rate provides a simple way to gauge the unwanted effects of the drive, with the coupler excitation providing an estimate for the expected coupler-induced infidelity of the process.

We simulate the coupler in its undriven eigenbasis, with the first 50 eigenstates.
We calculate the Floquet modes $\vert \alpha_{n} \rangle$, which for small amplitudes  adiabatically map to the $n$-th undriven eigenstate (Fock state $\vert n\rangle$).
We calculate the transition rates $W_{mn}$ from $\vert \alpha_{m} \rangle$ to $\vert \alpha_{n} \rangle$ via
\begin{linenomath*}
\begin{align}
    W_{mn} &= \sum_{k=-\infty}^{+\infty}S_{FF}\left[\Delta_{nmk}\right]\nonumber\\
    &\times \vert\int_{0}^T \frac{e^{-ik\omega_d t}}{T}\langle\alpha_m\left(t\right)\vert\hat{O}\vert\alpha_n\left(t\right)\rangle dt\vert^2.\label{FGR_Floquet}
\end{align}
\end{linenomath*}
Here, $\omega_d$ is the greatest common divisor of the frequencies of the two drives, $\omega_{d1}$ and $\omega_{d2}$, and $T=2\pi/\omega_d$ is the least common multiplier of their periods.
$\hat{O}$ (in the simulation taken to be the charge operator) is the system operator that is coupled to the bath, through $\hat{\mathcal{H}}_{int}/\hbar = \delta F\left(t\right)\hat{O}$, where the effect of bath is modelled by the noise parameter $\delta F\left(t\right)$.
The fluctuations in this parameter are captured by $S_{FF}\left[\omega\right]$, the noise spectral density:
\begin{linenomath*}
\begin{equation}
    S_{FF}\left[\omega\right] = \int_{-\infty}^{+\infty} d\tau e^{i\omega \tau}\left<\delta F\left(\tau\right)\delta F\left(0\right)\right>.
    \label[eq]{eq:spectral_density}
\end{equation}
\end{linenomath*}
The argument of this spectral density in Eq.~\ref{FGR_Floquet}, $\Delta_{nmk}=\omega_m - \omega_n - k\omega_d$, represents all the possible frequencies at which the transition of interest is coupled to the bath. 

We start with simulating the effects of dressed coupler dephasing, by looking at the lab-frame Hamiltonian of the coupler in the presence of both drive tones and incoherent coupler frequency fluctuations (\cref{floquet_sims}a).
We introduce a useful quantity, the drive asymmetry $l$,
\begin{linenomath*}
\begin{equation}
    l=\frac{\lvert\phi_c\rvert}{\sqrt{\lvert\phi_c\rvert^2+\lvert\phi_d\rvert^2}},
    \label[eq]{eq:meth_asymmetry}
\end{equation}
\end{linenomath*}
for which 0 corresponds to a fully differential drive, and 1 to a fully common mode drive. As a function of $l$, we calculate $W_{01}\times t_{\text{SWAP}}$, which measures the coupler excitation after a single Alice-Bob swap operation.
For a fair comparison, we also re-scale drive amplitudes to maintain a total beamsplitter rate of $g_{BS}/2\pi \approx 2$MHz, corresponding to $t_{\text{SWAP}}=125$ns.

We find, as predicted by theory, that this effect is primarily sensitive to the noise spectral density of coupler frequency fluctuations ($S_{\omega_c \omega_c}[\Omega]$) at frequencies close to  the detuning of the drive and the coupler mode ($\Omega \sim \Delta_{d_{1,2}, c}$).
In flux tunable devices, this noise spectrum is expected to decrease inversely with increasing frequency, suggesting that with a reasonable large drive detuning ($\sim 4.25$ GHz for this experiment), the effect of flux-noise induced dressed dephasing should be negligible.
We show here however, that that experimentally achievable drive symmetries suppress dressed dephasing by several orders of magnitude regardless of the dephasing rate (\cref{floquet_sims}a).

Through a similar Floquet-Markov calculation with both tones, we also calculate an effective coupler heating rate arising from a dressing of intrinsic coupler decay (\cref{floquet_sims}b), specifically for the differentially driven Hamiltonian.
Here, we find that despite the parity protection, there exists a significant heating rate within $\sim 1$ GHz of the resonant squeezing, which occurs near the coupler frequency.
This heating can intuitively be understood as a combination of a $\vert g \rangle \rightarrow \vert f \rangle$ coherent transition due to squeezing, and a $\vert f \rangle \rightarrow \vert e \rangle$ incoherent transition due to coupler decay.
Choosing a large drive-detuning strongly suppresses the squeezing strength and avoids this issue.
We also find weaker resonances at frequencies that are at a half ($\phi_d^4 {c^\dagger}^2$) and a third ($\phi_d^6 {c^\dagger}^2$) of the resonant squeezing process, but we are able to utilize the freedom in choosing our drive frequencies to evade them.

Finally, to provide an intuition for the combined effect of the odd-order processes forbidden by the parity protection, we show the marked difference in the allowed `drive space' for the common and the differentially driven SQUID (\cref{floquet_sims}c).
We define the drive space to be the allowed frequencies and amplitudes of a single drive tone that can drive the converter while keeping the coupler in its driven ground state.
Since a large portion of either plot shows chaotic behavior and coupler ionization \cite{TransmonIonization}, we use the impurity of the steady-state density matrix as our performance metric, which circumvents the need for a precise mapping between the Floquet states and the undriven states.
Since these are steady state calculations, the impurity of the coupler in regions where it is not ionized should not directly be interpreted as the heating rate for short time scales.
We observe that the differential drive's allowed frequency and amplitude range is much larger than those allowed for the common drive, clearly demonstrating the former's advantage in this simplistic scenario.
This simulation only contains a single drive tone and simulates just the coupler mode, but  we expect this advantage in drive space to hold even in the presence of the coupled storage modes and multiple drives, since a large fraction of the additional parasitic processes will also be forbidden for the parity-protected Hamiltonian.

\section{Suppressing the back-action of the buffer cavity}
While we have discussed the elimination of a linear coupling between the buffer mode and the coupler in our Methods, quantum fluctuations in the buffer mode might still dephase the coupler even in the presence of a purely non-linear coupling. This can be seen by writing the buffer-SQUID Hamiltonian in the dressed ladder-operator basis,
\begin{linenomath*}
\begin{align}
&\hat{\mathcal{H}}/\hbar=\omega_c \hat{c}^\dagger \hat{c} + \omega_B \hat{B}^\dagger \hat{B} \nonumber \\ &- \frac{E_{J_1}}{\hbar}\cos_{\textup{NL}}\left(\hat{\theta}_c+k\hat{\theta}_B\right)-\frac{E_{J_2}}{\hbar}\cos_{\textup{NL}}\left(\hat{\theta}_c-k\hat{\theta}_B\right)\nonumber\\
&\approx \omega_c \hat{c}^\dagger \hat{c} + \omega_B \hat{B}^\dagger \hat{B} - \chi_{Bc} \hat{B}^\dagger \hat{B} \hat{c}^\dagger \hat{c}.
\label{eq:coupled_SQUID_buffer_normal}
\end{align}
\end{linenomath*}
Here, $\omega_B$, $\hat{B}$ and $\hat{\theta}_B$ are the frequency, the ladder operator, and the phase operator of the buffer mode. $k=\frac{1}{2}M_{Bc} L_B^{-1}$ measures the participation of the buffer-mode phase in the superconducting phases across the SQUID junctions, with $L_B$ and $M_{Bc}$ being the self-inductance of the buffer mode and the mutual inductance between the buffer mode and the SQUID loop, respectively. The cross-Kerr term emerges from the differential coupling with a strength of $\chi_{Bc}\approx E_J \theta_{c,\text{zpf}}^2 \theta_{B,\text{zpf}}^2 k^2$, and translates the photon-number fluctuations in the buffer mode to coupler dephasing noise. For thermal-photon distribution and coherent-photon distribution, the dephasing rates are given by~\cite{Fei2018}
\begin{linenomath*}
\begin{align}
    \Gamma_{\phi}^{\text{th}}&=\chi_{Bc}^2 \frac{\overline{n}_{\text{th}}}{\kappa_B}, \nonumber\\
    \Gamma_{\phi}^{\text{coh}}&=\chi_{Bc}^2 \frac{\overline{n}_{\text{coh}}\frac{\kappa_B}{2}}{\left(\omega_d-\omega_B\right)^2+\left(\frac{\kappa_B}{2}\right)^2}, \label{eq:dephasing_rates}
\end{align}
\end{linenomath*}
where $\overline{n}_{\text{th}}$ and $\overline{n}_{\text{coh}}$ represent the mean thermal-photon number and coherent-photon number of the buffer mode, respectively. On the other hand, the beamsplitter rate is only linearly dependent on the cross-Kerr strength, $g_{BS}\propto \chi_{Bc} \overline{n}_{\text{coh}}$.
This allows us to suppress these dephasing rates while maintaining the beamsplitting rate, simply by reducing the cross-Kerr strength, and increasing the buffer drive strength (to increase $\overline{n}_{\text{coh}}$). We calculated using a HFSS-EPR simulation \cite{EPRPaper} that our design has $\chi_{Bc}\approx 2\pi\times 0.1$kHz. which corresponds to negligible dephasing rates of $\Gamma_{\phi}^{\text{th}}\ll 1$Hz, $\Gamma_{\phi}^{\text{coh}}\ll 1$kHz. With a drive power of $P\approx -50$dBm at the buffer cavity drive port, we can realize a differential drive amplitude of $\phi_d\approx 0.2\Phi_0$, that drives a sufficiently fast beamsplitter for this experiment.

\section{Decoherence in the dual-rail subspace}
In this section, we derive the evolution of the dual-rail density matrix (Methods Eq. 24), and discuss how the dual-rail decay and dephasing rates ($\kappa_1$ and $\kappa_{\varphi}$) are related to fluctuations in the undriven system.
Without loss of generality, the rotating frame Hamiltonian of the driven dual-rail qubit is given by
\begin{linenomath*}
\begin{align}
\hat{\mathcal{H}}(t)/\hbar&= \left(g_{\text{BS}}+\delta g(t)\right) \left(\hat{a}^\dagger \hat{b} +\hat{a}\hat{b}^\dagger\right) \nonumber\\
&+\delta\omega_a (t) \hat{a}^\dagger \hat{a} + \delta\omega_b (t) \hat{b}^\dagger \hat{b} \nonumber \\
&+\delta f_a (t)\hat{a} e^{-i\omega_a t}+\delta f_b (t)\hat{b} e^{-i\omega_b t}+ h.c. \label[eq]{eq:rot_Hamiltonian_with_g_noise}
\end{align}
\end{linenomath*}
Here, $\delta g$ represents the fluctuation of the beamsplitter rate, which could arise from a dispersion of beamsplitting strength with respect to beamsplitting strength.
$\delta\omega_{a,b}$ are the frequency fluctuations of the Alice and Bob cavities, which could arise from their cross-Kerr to the coupler mode, or from sources like a hot ancilla or other intrinsic dephasing.
Imperfections in control electronics, including slower drifts over time, could also contribute to these amplitude or frequency fluctuations.
Finally, $\delta f_{a,b}$ are the cavities' couplings to environmental decay channels.
These noises ($\delta x = \{\delta g, \delta \omega, \delta f\}$) are characterized by their spectral density (\cref{eq:spectral_density}), 
which are related to the decay and dephasing rates of the individual cavities,
\begin{linenomath*}
\begin{equation}
\kappa^{a,b}_{\downarrow}=S_{f_{a,b}f_{a,b}}[\omega_{a,b}], \;\kappa^{a,b}_{\phi}=S_{\omega_{a,b}\omega_{a,b}}[0]/2.
\end{equation}
\end{linenomath*}

To analyze how \cref{eq:rot_Hamiltonian_with_g_noise} gives rise to decoherence in the dual-rail subspace, we work in the eigenbasis of the beamsplitter interaction, $\lvert \text{Alice},\text{Bob}\rangle=\lvert00\rangle$, $\vert\psi^{\pm}\rangle=\left(\vert01\rangle \pm \vert10\rangle\right)/\sqrt{2}$, and $\lvert11\rangle$, and rewrite this Hamiltonian in the matrix form,

\begin{linenomath*}
\begin{align}
&\hat{\mathcal{H}}(t)/\hbar=\nonumber \\
&\begin{bmatrix}
0 & \delta f_- (t) & \delta f_+ (t)  & 0\\ 
 \delta f_- (t)^* & \delta\Omega_-(t) & \frac{\delta \omega_- (t)}{2}e^{-2ig_{\text{BS}} t}  & -\delta f_- (t) \\
\delta f_+ (t)^* & \frac{\delta \omega_- (t)^*}{2}e^{2ig_{\text{BS}} t} & \delta \Omega_+(t) & \delta f_+ (t) \\ 
0 & -\delta f_- (t)^* & \delta f_+ (t)^*  & \delta \omega_+(t)\label{eq:matrix_with_g_noise}
\end{bmatrix},
\end{align}
\end{linenomath*}
where
\begin{linenomath*}
\begin{align*}
\delta\Omega_{\pm} &= \delta \omega_a (t)/2+\delta \omega_b (t)/2\pm\delta g(t),\\
\delta f_{\pm} &= \left(\delta f_a (t) e^{-i\omega_a t} \pm \delta f_b (t) e^{-i\omega_b t}\right)\frac{e^{\mp ig_{\text{BS}} t}}{\sqrt{2}}, \\
\delta \omega_{\pm} &= \delta\omega_a (t) \pm \delta\omega_b (t).
\end{align*}
\end{linenomath*}
  This allows the calculation of the transition rates among the four eigenstates using time-dependent perturbation theory \cite{Clerk2010}, from which we obtain the decay rates from $\vert\psi^{\pm}\rangle$ to $\lvert00\rangle$ as $\kappa_{\psi^{\pm}}=\left\{S_{f_a f_a}[\omega_a\pm g_{\text{BS}}]+S_{f_b f_b}[\omega_b\pm g_{\text{BS}}]\right\}/2$. The relaxation between  $\lvert\psi_{\pm}\rangle$ due to the cavity frequency fluctuation $\delta\omega$, as well as the beamsplitter rate fluctuation $\delta g$, lead to dephasing in the dual-rail subspace at a rate $\kappa_{\varphi}=\left\{\bar{S}_{\omega_a \omega_a}[2g_{\text{BS}}]+\bar{S}_{\omega_b \omega_b}[2g_{\text{BS}}]\right\}/4+S_{gg}[0]$, with $\bar{S}\left[\omega\right]=\left\{S\left[\omega\right]+S\left[-\omega\right]\right\}/2$ being the symmetrized spectral density. At small $g_{\text{BS}}$ and weak dissipation rates, and assuming the cavity frequency fluctuations are uncorrelated, one can see how these rates are related back to cavity decay and dephasing rates, $\kappa_{\psi^{\pm}}=\kappa_1=(\kappa_{\downarrow}^a+\kappa_{\downarrow}^b)/2$, $\kappa_{\varphi}=(\kappa_{\phi}^a+\kappa_{\phi}^b)/2+S_{gg}[0]$. 

From here, it is straightforward to write down the evolution of the dual-rail qubit's density matrix, in the basis of $\lvert\psi^{\pm}\rangle$,
\begin{linenomath*}
\begin{align}
&\rho_{\vert\psi^{\pm}\rangle}(t) =\frac{1}{2}e^{-\kappa_1 t}
\begin{bmatrix}
1 & e^{i\Omega t-\kappa_\varphi t}\\ 
e^{-i\Omega t-\kappa_\varphi t} & 1
\end{bmatrix},
\label{eq:dm_dual_rail_supp}
\end{align}
\end{linenomath*}
where $\Omega = 2g_{\text{BS}}$. This can be easily transformed to  $\lvert10\rangle$ and $\lvert01\rangle$ basis (denoted as $\rho_{\text{DR}}$),
\begin{linenomath*}
\begin{align}
&\rho_{\text{DR}}(t) =\frac{1}{2}e^{-\kappa_1 t}\times\nonumber\\
&
\begin{bmatrix}
1+e^{-\kappa_{\varphi}t}\cos\left(\Omega t\right) & i\sin\left(\Omega t\right) e^{-\kappa_\varphi t}\\ 
-i\sin\left(\Omega t\right) e^{-\kappa_\varphi t} & 1-e^{-\kappa_{\varphi}t}\cos\left(\Omega t\right)
\end{bmatrix},
\label{eq:dm_dual_rail_01basis}
\end{align}
\end{linenomath*}
which becomes Eq. 24 in Methods.

\begin{figure*}[!ht]
\centering
\includegraphics[width=0.98 \textwidth]{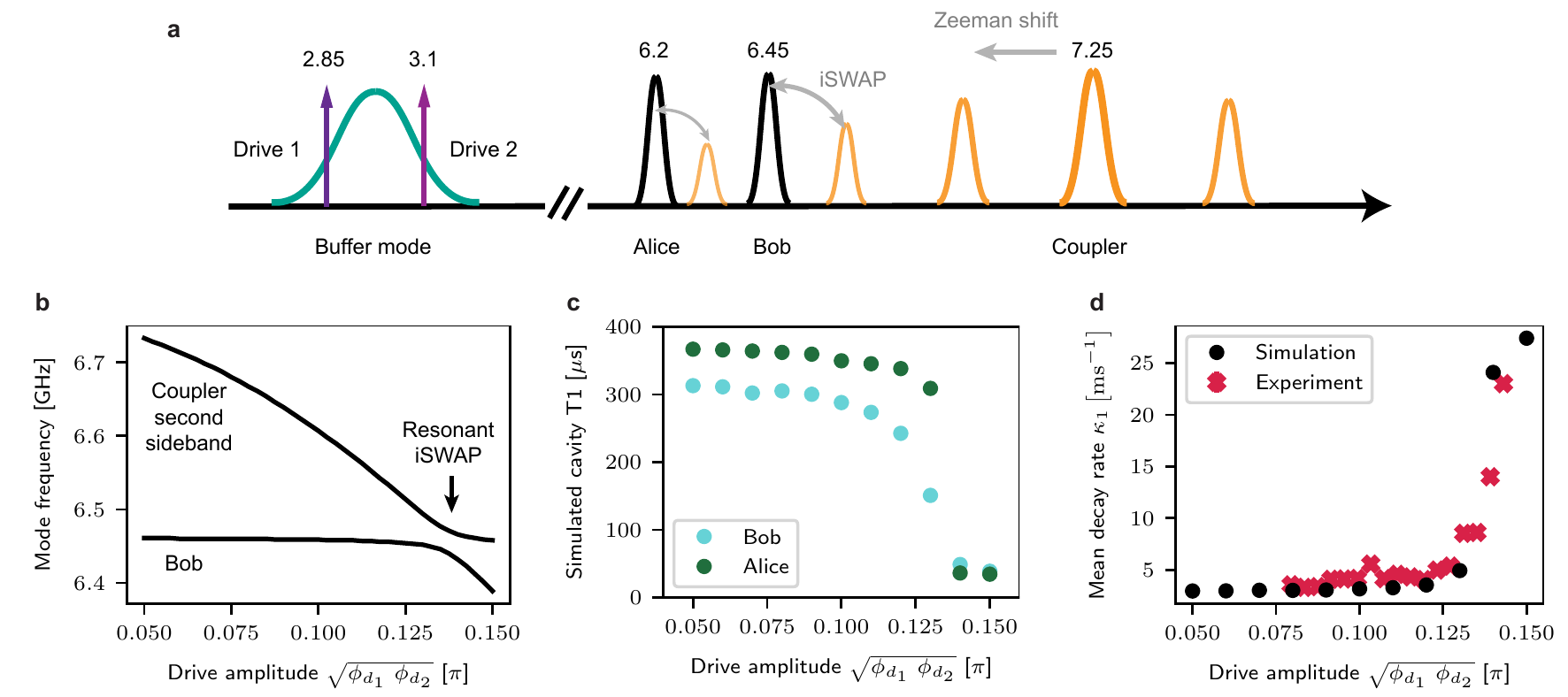}
\caption{
\textbf{Coupler-cavity sideband collision on increasing drive amplitude}
\textbf{a}, Frequency stack of the driven system. Due to the Zeeman shift, the coupler frequency is pushed towards the Alice and Bob cavities.
The parametric modulation also creates sidebands of the coupler that are spaced by the two-tone detuning.
These sidebands are coupled to the cavities through detuned iSWAP interactions, that get closer to resonance at stronger drive amplitudes.
\textbf{b}, Simulation of the frequencies of the Bob cavity and the second (red-detuned) sideband of the coupler as a function of the drive amplitude. A clear avoided crossing is seen around drive amplitude of 0.14$\pi$ when the coupler sideband collides into Bob, corresponding to a resonant iSWAP interaction.
\textbf{c}, The Floquet-Markov simulations showing the reduction of the cavities' lifetimes with respect to the drive amplitude. 
Alice's and Bob's $T_1$s are obtained from two separate simulations, where only Alice or Bob mode is included along with the coupler, respectively. In both simulations, decay channels were set to produce similar undriven lifetimes of Alice, Bob and the coupler to the experimental values.
\textbf{d}, A comparison of the mean of the decay rates (black circles) obtained from the simulation in \textbf{c} to the experimentally obtained decay rate of the dual-rail qubit (red crosses, same as main text Fig. 7).
The curves show qualitative agreement, but the simulations only capture processes that involve the individual cavities and the coupler, and therefore a quantitative agreement is not expected.
}\label[fig]{sideband_collision}
\end{figure*}

\section{Limits on performance from coupler-storage sideband collision}
We now provide a hypothesis for the limitations on fidelity and achievable beamsplitting strength that we find in our device. We see in experiment (Methods Fig. 7) that the effective beamsplitter decoherence rate remains stable as a function of drive amplitude, until it suddenly rises at $\vert \phi_d \vert \sim 0.13\pi$.
This is accompanied by a super-quadratic dependence of $g_{\text{BS}}$ on the drive amplitude.
A hypothesis for this effect is an inflation of the interaction between the cavities and the coupler, mediated by the beamsplitter drives.
This could arise from a drive-induced coupler-cavity sideband interaction, that is then magnified by the coupler's Zeeman shift in the regime of strong drives.
Here, we investigate the strength of such a 'sideband collision', first focusing on Bob, which is closer in frequency to the coupler.
By combining Eqs. 11,12, and 15 in the main text, one obtains a photon exchange interaction between the coupler and Bob in the rotating frame:
\begin{linenomath*}
    \begin{align}
        \frac{\hat{\mathcal{H}}_{bc}^{(\text{RWA})}}{\hbar} &\approx \sum^\infty_{n=1} \omega_c J_n\left(\vert\phi_{d_1}\vert\right)J_n\left(\vert\phi_{d_2}\vert\right)\cos\left(n\Delta_d t\right)\nonumber\\
        &\times \beta_b \left(e^{i\Delta_{bc}t} \; \hat{b}^\dagger \hat{c}+\text{h.c.}\right).
        \label[eq]{eq:sideband_interaction}
    \end{align}
\end{linenomath*}
Here, $\Delta_{bc} = \omega_b-\omega_c+\omega_{\text{Z},b}-\omega_{\text{Z},c}$ represents the detuning between Bob and coupler after accounting for Zeeman shifts.
When the modulation frequency satisfies $\Delta_d=\Delta_{bc}/n$, a resonant `iSWAP' interaction occurs between Bob and the coupler, which we interpret as the n-th sideband collision. 
This can be intuitively understood as a sideband interaction~\cite{FirstOrderSideband}, as illustrated in \cref{sideband_collision}a.
The two-tone flux modulation creates coupler sidebands spaced by the tone detuning $\Delta_d$ which, when pushed down by the Zeeman shift, start interacting strongly with the storage cavities.
We identify from Floquet simulations that around a drive amplitude of $0.14\pi$, the second sideband of the coupler crosses the Bob mode (\cref{sideband_collision}b), thereby limiting Bob's lifetime due to an inherited decay.
We expect Alice to show similar behaviour, encountering the coupler's third sideband at a similar Zeeman shift, but we expect this process to be higher order and therefore weaker.

To study the implications of this increased hybridization, we perform Floquet-Markov simulations to find predicted $T_1$'s for both cavities as a function of drive strength (\cref{sideband_collision}c).
These decay rates are extracted from two separate simulations, with each simulation including the coupler mode and only a single cavity mode, but should be sufficient to capture the iSWAP interactions.
We therefore do not necessarily expect the simulations to show quantitative agreement with experiment, as the experiment could be limited by processes that simultaneously involve both cavities, or the coupler's readout mode.
However, for a qualitative test of this hypothesis, we compare the predicted mean cavity decay rates $\kappa_1$ to those found in experiment (main text Fig. 7), and find a reasonable agreement (\cref{sideband_collision}d).
This issue of increased hybridization could potentially be avoided by redesigning the coupler driven frequency at the operating point, for example by placing it red-detuned to the cavities, such that the cavities are limited by a higher-order sideband.

\section{Cavity nonlinearity induced by the driven SQUID}

\begin{figure*}[ht]
\centering
\includegraphics[width=0.98\textwidth]{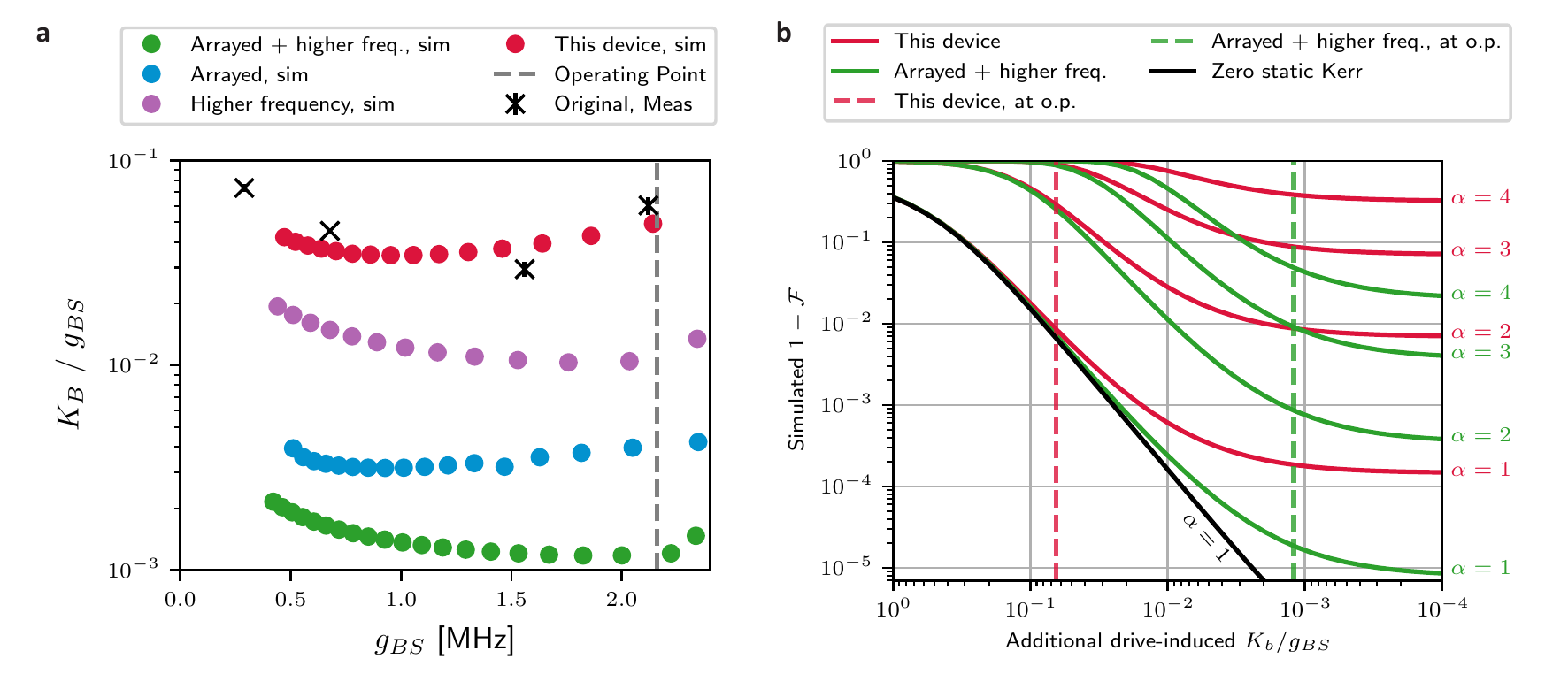}
\caption{ \textbf{Drive-induced cavity Kerr.}
\textbf{a},The ratio between the beamsplitter rate, $g_{BS}$, and the self-Kerr of Bob, $K_B$, plotted against the beamsplitter rate. In the original circuit (red dots and blue crosses), this ratio is limited by the Zeeman shift as well as the sideband collision issue. By rearranging the frequency stack such that the coupler is further detuned from the cavities ($f_c=$7.75GHz, adjusting $g_{ac}$ and $g_{bc}$ accordingly to maintain the coupler-cavity hybridization), this ratio increases by almost a factor of 5 (orange dots), at the operating point where $g_{BS}/2\pi\approx 2$MHz. Further optimization is possible by arraying the device into 3 SQUIDs (green dots), which roughly increases this ratio by a factor of $3^2=9$. \textbf{b}, Swap operation infidelity for various coherent states as a function of drive-induced Kerr, at fixed $g_{BS}/2\pi = 2.16$MHz. Static Kerr is set to be the same as the current device (red), and the arrayed device with 3 SQUID (blue). Cavity Kerr leads to infidelity through both an undesired photon-number-dependent phase accumulation as well as a shift in the beamsplitting resonance condition for each photon manifold. When $K_b/g_{BS} \sim 1$, this amounts to photon exchange only occurring in a single photon manifold. As $K_b/g_{BS}$ is reduced the swap fidelity of coherent states reduces quadratically with this ratio (highlighted by the black line corresponding to zero static Kerr), until it saturates to an infidelity limit imposed by $K_b^{\text{static}}/g_{BS}$. The red and green dashed lines indicate the drive-induced $K_b/g_{BS}$ of the current device, and the arrayed SQUID at higher frequency, respectively.}
\label[fig]{fig:drive_induced_Kerr}
\end{figure*}

In addition to the increased cavity decay rates due to the sideband collision, large drive amplitudes on the SQUID coupler may also induce more nonlinearity to the cavities.
We investigate the magnitude of this induced nonlinearity, it's effect on the beamsplitter, and strategies to mitigate the driven nonlinearity in this section.

A natural contribution to the static Kerr of the cavities comes from the SQUID coupler, which is in the transmon regime and has a self-Kerr of $K_c/2\pi\approx125$MHz.
As a result of the linear dispersive coupling, the storage cavities weakly hybridize with the coupler and inherit self-Kerrs, $K_{a(b)}\approx\left(\frac{g_{a(b)c}}{\Delta_{a(b)c}}\right)^4K_c$. 
We directly measure this self-Kerr by first displacing the cavity to $\vert\alpha\rangle$, followed by a delay time of $t$ that incurs a phase space rotation proportional to $ \Delta+\alpha^2\frac{K_b}{2}$, where $\Delta$ is the detuning between the cavity and drive frequencies.
We then use a second displacement pulse with opposite phase that attempts to bring the cavity back to vacuum, with full revival determined by $\Delta$, $K_b$ and $t$.
By performing a measurement conditioned on the cavity being in the vacuum state using the ancilla qubit and repeating this measurement as a function of $\alpha$, we are able to quantify this revival and extract both the Kerr and detuning~\cite{SNAPPaperWithKerrMeasurement}.

The cavity self-Kerrs are measured to be $K_a/2\pi\approx 5$kHz and $K_b/2\pi\approx 15$kHz when the coupler is undriven. When the coupler is driven, the Zeeman shift pushes the coupler frequency lower, bringing it closer to the cavities and therefore imposing a higher amount of nonlinearity to the cavities.
This is accompanied by the sideband collision discussed in the previous section, which also causes a stronger hybridization between the coupler and the cavities. At the operating point where a beamsplitter rate of $\sim2.2$MHz is achieved, we measure the self-Kerr of Bob cavity to be around $128$kHz.
We expect the self-Kerr of Alice to remain lower than Bob's, since the sideband it encounters is higher-order.
While this aggravated nonlinearity is relatively harmless in the dual-rail subspace central to our investigation, it could introduce a coherent error in beamsplitter operations that involve large photon numbers, and optimizing the DDS' performance by reducing the drive-induced Kerr is an important future direction.

There exist multiple solutions that are fully compatible with our architecture, as demonstrated in \cref{fig:drive_induced_Kerr}. For instance, since the dominant source of cavity nonlinearity is from the Zeeman shift and the sideband collision, we could rearrange our mode frequencies such that the coupler is detuned further from the cavities. Thus, at the same drive amplitude (that produces the same $g_{BS}$), both the coupler and its sideband stay further away from the cavities, which is helpful for decreasing the Kerr$/g_{BS}$ ratio. We could also redesign the coupler as an array of dc SQUIDs, which would suppress the coupler Kerr by a factor $N^2$, where $N$ is the number of the SQUID loops. Finally, the cavity Kerrs can also be dynamically canceled through additional microwave drives on the coupler, as has been shown in \cite{Yaxing_Kerr}, or corrected via additional unitaries like SNAP \cite{SNAPPaperWithKerrMeasurement}.

To place the quantity $g_{BS}/K_b$ into context, 
and to get a sense of how significantly this nonlinearity affects higher-photon manifolds,
we examine the limits on swap fidelity imposed by this coherent error.
Using time-domain QuTiP \cite{qutip} simulations, we first calibrate a swap pulse at our operating point of $g_{BS}/2\pi = 2.16$ MHz optimized for the single-photon manifold, as done in experiment. 
We then apply this pulse to a coherent state in the presence of both static and drive-induced Kerr (scaling all driven Kerrs proportionally), computing the final state fidelity 
$\vert\langle \psi_{ideal} \vert \psi_{final} \rangle\vert^2 $ to the ideal swapped coherent state $\vert \psi_{ideal} \rangle$.

\cref{fig:drive_induced_Kerr}b shows this swap infidelity as a function of drive-induced $K_b/g_{BS}$ for different coherent states and static Kerrs. When the drive-induced Kerr is on the order of the beamsplitting rate, the coherent state swap fidelity suffers drastically as the beamsplitter becomes increasingly off-resonant for all but the single-photon manifold. As the driven Kerr is reduced, the infidelity reduces quadratically for decreasing $K_b/g_{BS}$ until eventually saturating to a background infidelity limit imposed by the static Kerr.

These simulations demonstrate that by implementing modifications compatible with our design, we can achieve a substantial improvement in performance. Arraying the coupler ($N = 3$) and shifting to a higher frequency would lead to a Kerr-limited swap fidelity of more than $99.9\%$ for the $\vert \alpha = 2\rangle$ coherent state, and above $99\%$ for $\vert \alpha = 3\rangle$. Note that this effect could be mitigated to some extent by calibrating the beamsplitter in photon manifold corresponding to the mean coherent state photon number. While applications with much more stringent linearity requirements exist \cite{GKPExperiment,VladGKPBeyondBreakeven}, significant progress towards realizing hardware-efficient error correction has been made with protocols using states with mean photon numbers as low as the states simulated here \cite{BeyondBreakeven2016,ChenWangAQEC,BinomialBeyondBreakeven}.
However for most applications to high fidelity control of logical qubits, including an additional unitary resolving this coherent error will be useful.

\section{DC flux calibration and junction asymmetry}
\begin{figure*}[ht]
\centering
\includegraphics[width=0.98\textwidth]{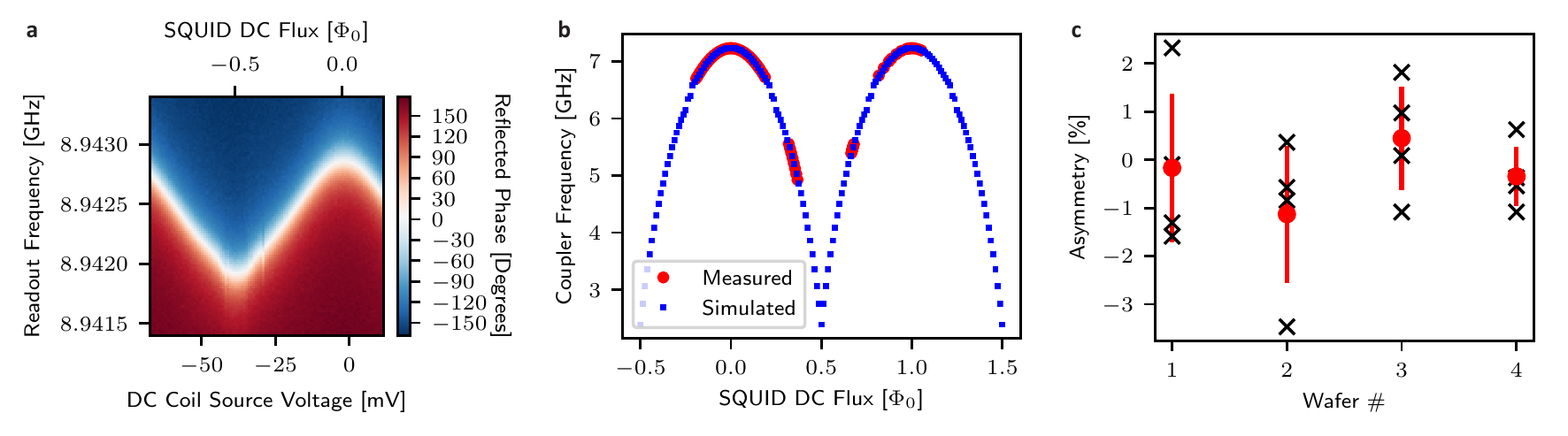}
\caption{ \textbf{DC flux and junction asymmetry.}
\textbf{a,} Spectroscopy of the coupler readout resonator while tuning source voltage of the DC flux coil. This curve provides an initial estimate of the zero flux point as well as the conversion between coil voltage and Flux Quantum.
\textbf{b,} Coupler frequency vs. DC flux. The data is fit to a model that takes both linear inductance of the SQUID loop and hybridization between coupler, Alice, and Bob into account.
\textbf{c, } Junction asymmetry as backed out from room temperature resistance measurements from SQUID devices on multiple wafers. These devices were fabricated using the same methods as (and contemporaneous with) the device demonstrated in the main text (\cref{fab_notes}).
}\label[fig]{flux_and_LJ}
\end{figure*}

Operating at the DC flux sweet spot of the SQUID is not only desirable for combating flux noise, but is also critical to keeping our aforementioned parity-protection.
To ensure that we operate at $\Phi_{\text{DC}} = 0$, we have a coil built in to our package to apply a static magnetic field through the SQUID loop. This coil is made out of thin NbTi wire wrapped around a copper spool, similar to the magnet in many resonant JPA designs, and is placed sufficiently far from the converter and cavities, so as to not spoil any high-Q modes. The coil is connected to a DC power supply, where we can tune the source Voltage as desired.

The spectroscopy of the coupler readout serves as an initial estimate of the zero flux point, where we observe a periodic tuning of the coupler readout frequency as the DC source is swept, due to the repulsion of the coupler mode.  (\cref{flux_and_LJ}a).

To obtain a more precise and more direct measurement of the zero-flux point, we then perform coupler spectroscopy while sweeping the now-calibrated DC source (\cref{flux_and_LJ}b).
The coupler frequency is measured near zero flux as well as further on the slope.
As the coupler frequency sweeps into resonance with Alice and Bob, the modes strongly hybridize, obscuring the coupler resonance.
The measured coupler frequency is well fit to a model that takes both the linear inductance in the SQUID loop and the hybridization with Alice and Bob into account, yielding the simulated points in \cref{flux_and_LJ}b.

Another requirement of the differentially-driven Hamiltonian (Eq. 1 in the main text) is that the Josephson energies of the two junctions in the SQUID are the same, which in the practical sense means the junctions should have identical size. While this is the case in design, imperfections in the fabrication processes, especially in the e-beam writing step, results in an inevitable deviation of the $E_J$ of the two junctions. To quantify the typical value of the $E_J$ asymmetry, we measure the room-temperature resistance of the Josephson junctions, which is related to their Josephson energies through the Ambegaokar-Baratoff relation \cite{AB_relation}. Through this way, we measured the $E_J$ asymmetry, which we define as $\left(E_{J_1}-E_{J_2}\right)/\left(E_{J_1}+E_{J_2}\right)$, for the SQUID devices on four different wafers that were fabricated around the same time (\cref{flux_and_LJ}c). From these measurements, We observe a negligibly small $E_J$ asymmetry of 0.3\%$\pm$1.3\%. 

\section{Calibrating a beamsplitter pulse}
\label[Methods]{MethodsPulseCalibration}
In order to calibrate the beamsplitter that is used in the RB experiment, we must accurately tune up the pulses on the RF controls.
While fitting a chevron (Main text Fig. 4b) can approximately find the resonance condition, calibrating the system to the degree required for fidelities above 99.9\% requires more precise protocols.  
Using a four-wave mixing process to engineer the beamsplitter interaction also incurs drive-induced Zeeman shifts of the cavity frequencies that are on the same order as the beamsplitter rate, further complicating the calibration.
If these shifts were identical for Alice and Bob, then the beamsplitting resonance condition would not change, but any relative difference between these shifts will manifest as a shift in the beamsplitting resonance condition. 
Most notably, this means that the resonance condition changes as a function of drive amplitude, complicating the process of calibrating a pulse with finite ramps.

\begin{figure*}[ht]
\centering
\includegraphics[width=0.98 \textwidth]{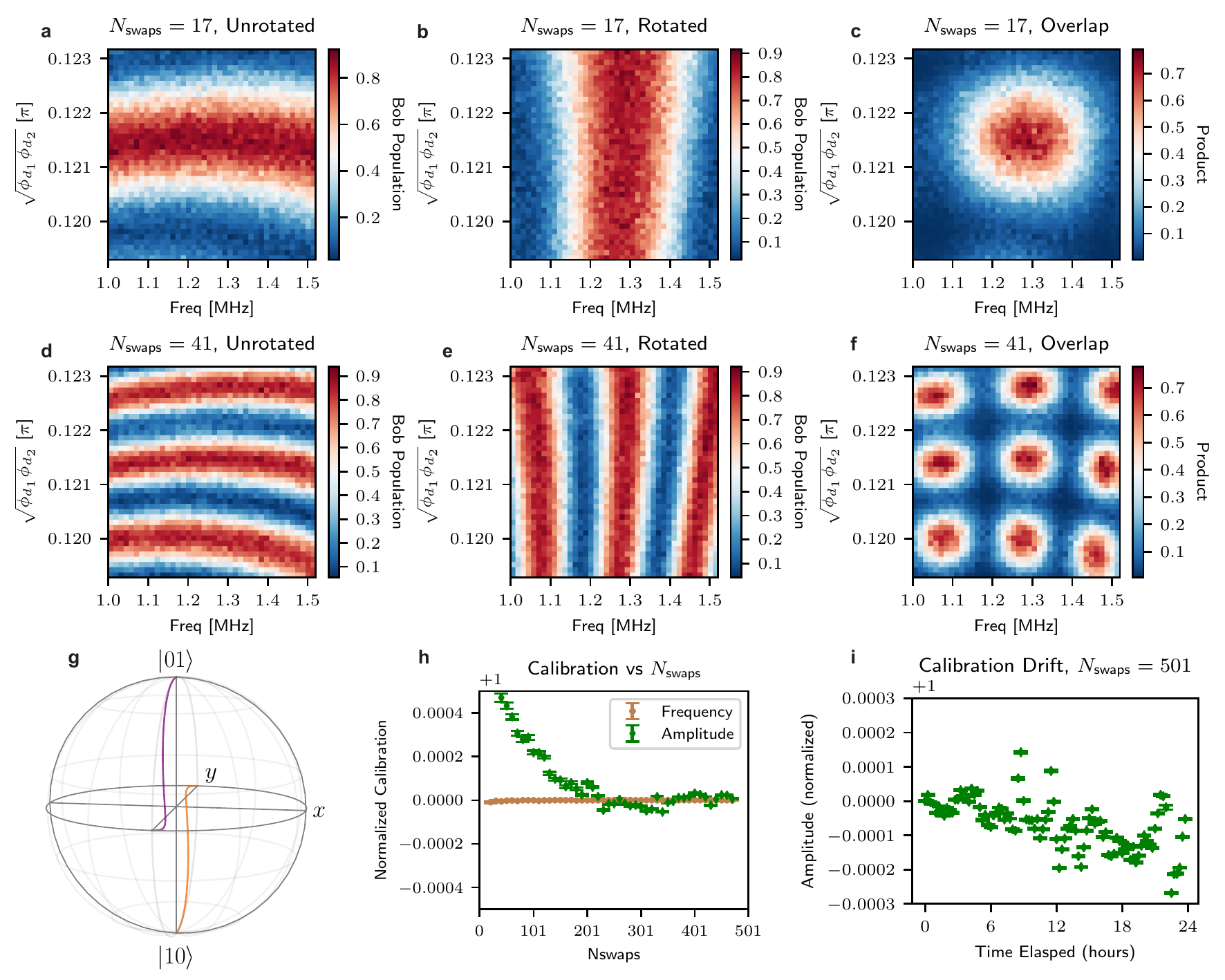}
\caption{
\textbf{Beamsplitter Pulse Calibration.} 
\textbf{a}, Sweeping drive amplitude and detuning, we perform $2N_{\text{swaps}}$ consecutive beamsplitter pulses with identical phase, leading to bands that correspond to plausibly correct calibration values.
\textbf{b}, By performing the same number of beamsplitter pulses as (a), but changing the phase of every two beamsplitters relative to the prior two, we can effectively engineer a sequence equivalent to $U_{\text{SWAP}} Z(\varphi) U_{\text{SWAP}} \dots Z(\varphi) U_{\text{SWAP}}$. For a value of $\theta_R = 1.3$ radians (empirically found), we recover bands that now trace paths roughly perpendicular to those found from the (\textbf{a}) protocol.
\textbf{c}, The correct calibration setting would pass both protocols, so we take the overlap, or product, of (\textbf{a}) and (\textbf{b}) to recover a region that the correct calibration is in.
\textbf{\textbf{d-f},} By increasing $N_{\text{swaps}}$, the bands get tighter individually, but also more densely spaced. This leads to multiple regions in the overlap plot (\textbf{f}), but only one region has the same center as regardless of $N_{\text{swaps}}$, in this case the center region. By iteratively narrowing the scope and increasing $N_{\text{swaps}}$, we can refine our calibration ever finer.
\textbf{\textbf{g},} Trajectories of states initialized on the poles of the dual-rail Bloch Sphere for the correct calibration. The pulse we calibrate for is detuned from the coherent evolution resonance condition, but the detuning combines with Zeeman shifts during the ramps (which appear as Z rotations on the Bloch sphere) to perfectly compensate and give the desired unitary.
\textbf{\textbf{h},} Calibration vs $N_{\text{swaps}}$. At each value of $N_{\text{swaps}}$, we fit the overlap region to a 2D Gaussian and extract center frequency (brown) and amplitude (green) as well as the fit error. By dividing these values by the final drive tone detuning ($\Delta_d = 237.26367 \text{ MHz} \pm 50 \text{ Hz}$) and final amplitude ($\sqrt{\phi_{d_1}\phi_{d_2}} = 0.1214367 \pm 4\times10^{-7}$) respectively, we can plot the calibration normalized to the final parameter set as a function of $N_{\text{swaps}}$
\textbf{\textbf{i},} To track the significance of calibration drifts, we run the protocol for $N_{\text{swaps}}=501$ every 15 minutes over a 24 hour period, finding no more than a 0.03\% fluctuation over this timescale. The frequency calibration has no observable drifts.
}\label[fig]{Pulse_Cal_Fig}
\end{figure*}

Within the dual-rail qubit, Eq. 18 in the main text reduces to
\begin{linenomath*}
    \begin{align}
        \hat{\mathcal{H}}_\text{BS}/\hbar &= ((\Delta_{ab} - \Delta_d) + \Delta_{\text{Z},ab} \varepsilon(t))\frac{\hat{\sigma}_Z}{2}  \nonumber \\
        &+ g_\text{BS}\;\varepsilon(t)\; (e^{i\varphi_\text{BS}} \hat{\sigma}_+ + e^{-i \varphi_\text{BS}}\hat{\sigma}_-),  \nonumber
    \end{align}
\end{linenomath*}
where $\varepsilon(t) \in [0,1]$ is a normalized envelope function corresponding to the square of the pulse shape, as both pulses are ramped simultaneously.

In this Bloch sphere picture, the polar angle $\theta(t)$ of a state starting at the North Pole is given by
\begin{linenomath*}
    \begin{align}
        \theta(t) &=  2\int_0^t g_\text{BS} \; \varepsilon(t') dt'. \nonumber
    \end{align}
\end{linenomath*}
The azimuthal phase drift during the operation is then given by
\begin{linenomath*}
    \begin{align}
        \delta\phi(t) &=  \int_0^t \sin(\theta(t')) \frac{((\Delta_{ab} - \Delta_d) + \Delta_{\text{Z},ab} \; \varepsilon(t'))} {2} \; dt'  \label[eq]{eq:phase_drift}
    \end{align}
\end{linenomath*}
There exists a static detuning $\Delta_d$ such that, at a time $\tau_\text{BS}$ given by $\theta(\tau_\text{BS}) = \frac{\pi}{2}$ we can have complete cancellation of the azimuthal phase drift $\delta\phi(\tau_\text{BS}) = 0$. Such a pulse takes the trajectory of a detuned process during the steady state before arriving back to the correct final position on the Bloch sphere during the ramp down  (\cref{Pulse_Cal_Fig}g).

To find the correct parameter set that achieves this, we can prepare a single photon in Bob, apply a candidate beamsplitter pulse a total of $4n+2$ times, and measure P(Bob = 0) to see whether the photon is successfully swapped out of Bob.
Specifically, we fix the total pulse time and envelope shape while simultaneously sweeping tone 1 drive frequency (and thus $\Delta_d$) and drive strength (and thus $g_\text{BS}$ and $\Delta_{\text{Z},ab}$) of the candidate pulse to scan the parameter space. If the candidate pulse does indeed implement a beamsplitter, then we will have implemented a total of $N_\text{swaps} = 2n+1$ swap operations on the photon, leaving Alice with the photon and Bob in vacuum.

To gain some insight as to what we should expect when sweeping these parameters, we consider an adapted Rabi model with an amplitude-dependent detuning. Under this model, we find the probability $P_\text{ B$\rightarrow$A}$ to oscillate as:
\begin{linenomath*}
    \begin{align}
        P_\text{ B$\rightarrow$A} &\approx \sin^2\left(2\sqrt{\Omega^2 + (\Delta_0 + \alpha \Omega)^2}t\right), \nonumber
    \end{align}
\end{linenomath*}
where $\Omega$ and $\Delta_0$ are drive strength and detuning respectively, and $\alpha$ is some coefficient relating drive strength to frequency shift. For a fixed time $T$, we would expect contours of constant $P_\text{ B$\rightarrow$A}$ to be of the form:
\begin{linenomath*}
    \begin{align}
        \Omega^2 + (\Delta_0 + \alpha \Omega)^2 &= \left(\frac{(2n+1)\pi}{4T}\right)^2. \nonumber
    \end{align}
\end{linenomath*}
This equation, when varying $\Omega$ and $\Delta_0$, results in ellipses with different bands corresponding to different values of $n$.
As seen in \cref{Pulse_Cal_Fig}a,d, a sweep of the parameter space does indeed recover these bands, with different bands corresponding to different odd integer fractions of the correct band. For example, a sequence that used $(4(5)+2)$ pulses but only implemented 3 total swaps would produce a band at $\frac{3}{5}$ of the radius of the correct band.

To narrow down precisely where along these bands the correct detuning to properly null ~\cref{eq:phase_drift} is, we run a complementary pulse sequence. This complementary sequence consists of pairs of candidate beamsplitter pulses, with the phase $\varphi_\text{BS}$ of each pair shifted by an angle $\theta_R$ relative to the previous pair. This relative angle between pulse pairs results in a rotation of the bands as shown in \cref{Pulse_Cal_Fig}b,e, and a value of $\theta_R = 1.3$ radians makes the new bands approximately orthogonal to the unrotated version. The correct set of calibration parameters successfully swaps the photon under both of these pulse sequences, so by taking the overlap (product) of these 2D sweeps we can get patches of possible parameter sets as seen in \cref{Pulse_Cal_Fig}c,f. 

Since the density of the contour lines is set by the number of pulses in the sequence, increasing $N_\text{swaps}$ not only makes the bands thinner individually, but also makes them appear closer together. Putting all of this together, we can start at a low value of $N_\text{swaps}$ where the bands are spaced far apart and the overlap region is wide. Then, by refining the parameter range and increasing $N_\text{swaps}$, we can iteratively make the overlap region smaller and tighten the bounds on the correct amplitude and detuning. Since this scheme amounts to detecting the center of a circle in an image, it is amenable to basic image processing techniques, allowing an automation of the calibration procedure. We use this to calibrate the beamsplitter pulse up to $\sim$1000 operations, limited by FPGA wave memory, leading to a fractional calibration precision of less than $3\times10^{-6}$ in Amplitude and $3\times10^{-7}$ in frequency (\cref{Pulse_Cal_Fig}h).

To monitor calibration drifts, the protocol was performed at $N_\text{swaps} = 501$ at 15 minute intervals for 24 hours (\cref{Pulse_Cal_Fig}i). We observe no significant drift in frequency, while observing a drift of no more than 0.03\% in amplitude. This amplitude drift, plausibly due to temperature fluctuations of the IQ-mixers and other control electronics, would only have a quadratic effect on the infidelity.

\section{Randomized Benchmarking protocols}
\label{RB_details}
Randomized benchmarking on a qubit encoded in a higher dimensional system presents unique challenges, particularly when leakage out of the qubit subspace is the dominant error \cite{LeakageRB}.
However, the dual-rail qubit \cite{JamesDualRailPaper} is still amenable to RB-like techniques because this leakage to vacuum can predominantly be detected and selected out.

To carry out a randomized benchmarking protocol on the dual-rail qubit, we construct the gate set $G_{\text{DR}}$ with the following mapping:
\begin{linenomath*}
\begin{align*}
    &\{X_{\pi/2}, Y_{\pi/2}, X_{-\pi/2}, Y_{-\pi/2}\} \rightarrow \\
    & \{U_{\text{BS}}(0), U_{\text{BS}}(\pi/2), U_{\text{BS}}(\pi), U_{\text{BS}}(-\pi/2)\}, \\
    \\
    &\{X_\pi, Y_\pi\} \rightarrow \{U_{\text{BS}}^2(0), U_{\text{BS}}^2(\pi/2)\}.
\end{align*}
\end{linenomath*}
Here, all $U_{\text{BS}}(\varphi)$ are generated by the same calibrated pulse (Methods 6.6), with varying drive phases (main text, Fig 3d).
Only the relative difference between the pulse phases matter, which the FPGA-based control can guarantee with high precision, and we expect this phase's accuracy to be significantly higher than that of the amplitude and detuning of the pulse.
We also purposefully exclude explicit $Z$ gates from our gate set, as these gates could be achieved by a phase update on subsequent gates and are independent of the beamsplitter fidelity.

\begin{figure*}[ht]
\centering
\includegraphics[width=0.98\textwidth]{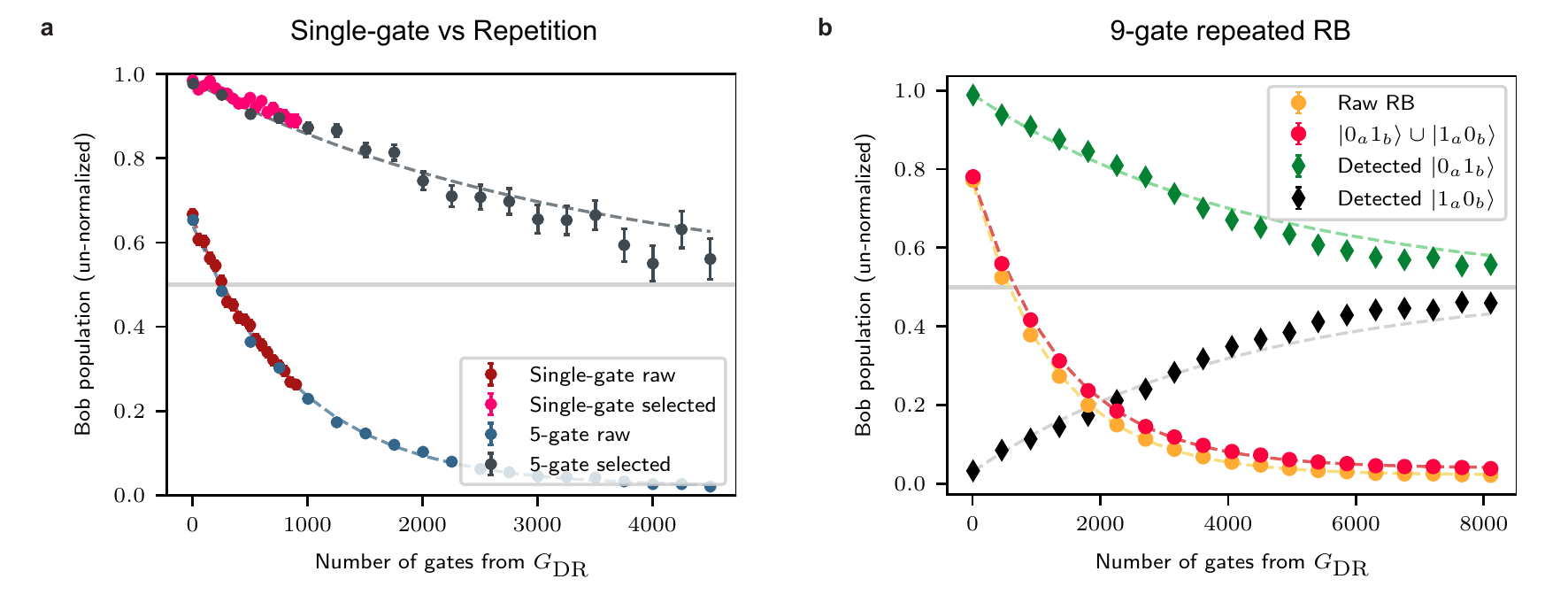}
\caption{ \textbf{Randomized Benchmarking protocols.}
\textbf{a,} Comparison of a standard RB protocol (each gate random) to a protocol with random gates repeated in sets of five.
The un-selected (raw) data agree within error-bars.
On post-selecting out single-photon loss events, the five-gate sequence provides a lower bound for the fidelity of the single-gate sequence, since repeating sets of gates amplifies the effects of coherent errors, like those due to miscalibration.
\textbf{b,} We analyze in detail a single RB experiment with random sets of 9 repeated gates, under various selection protocols, with no normalization. Dashed lines for each curve represent exponential fits.
The raw RB (yellow) shows a decay constant of $1271 \pm 4$ gates, with its reduced amplitude corresponding to a SPAM error of $\sim 21\%$.
The leakage-limited curve (red) represents the total rate of photon loss out of the dual-rail subspace, and does not distinguish between states on the dual-rail Bloch sphere. This curve has large overlap with the raw RB, implying that we are limited by photon loss errors, and shows a decay constant of $1388 \pm 4$ gates.
On detecting and selecting out the leakage events, we obtain error-detected curves where the system ended in either $\vert 0_a 1_b\rangle$ (green) or $\vert 1_a 0_b\rangle$ (black).
These sequences on average represent a depolarization channel, and seem to decay towards a perfectly mixed state (0.5, grey horizontal line) with improved decay times of $4477 \pm 30$ and $4194 \pm 32$ gates respectively.
All exponential fits take standard errors (SE) of each dataset into account, with data-points with fewer averages contributing less to the fit, which is particularly visible in the error-detected datasets.
}\label[fig]{fig:all_rb}
\end{figure*}

We generate sequences of up to 8100 such gates, by picking up to 900 random gates from $G_{\text{DR}}$ in real time and repeating each gate 9 times consecutively. The impetus for this unique choice of concatenation is due to an FPGA memory limit of $\sim 900$ consecutive random operations. By re-defining one `operation' to be $4n+1$ concatenations of the same gate (utilizing the fact that $g^{4n+1} = g, \forall g \in G_{\text{DR}}, \forall n \in \mathbb{N}$) we can go beyond this limitation to capture longer timescales.
At each sequence length, we generate $10^5$ such semi-random sequences for adequate precision even after post-selection.
These gates form an over-complete set of generators for the Clifford group, which can be seen by constructing the dual-rail Hadamard and $S$ gates:
\begin{linenomath*}
\begin{align}
    H_{\text{DR}} &= X_\pi Y_{\pi/2} \\
    S_{\text{DR}} &= Y_{-\pi/2} X_{\pi/2} Y_{\pi/2}.
\end{align}
\end{linenomath*}
Each sequence maps the initial state $\vert 0_a1_b \rangle$ to a known cardinal point on the Bloch sphere, after which we apply a single additional gate $U^{-1}(\varphi_n)$ that brings the state back to $\vert 0_a1_b \rangle$.
This state-mapping inverse differs from a full unitary inverse by up to a $Z$ rotation, but since the cavity measurements correspond to a logical Z measurement of the dual-rail qubit, the outcomes are unaffected by this difference.

This protocol is similar to `Direct RB' \cite{DirectRB}, with three notable exceptions - the leakage out of the subspace due to single-photon loss errors, the repeated gates, and the difference in the choice of the final Unitary.
Since the leakage is a detectable error, it can be separately selected out and quantified, leaving sequences that are fully randomized to a depolarization channel under a sufficiently large number of averages and sufficiently deep circuit ($\sim 10^4$ averages for the first $2250$ gates), assuming effects due to a difference in the cavity decay rates is negligible (see \cref{coherences} for coherences).
We expect the repeated gates to slightly enhance the effect of systematic errors, like those in our calibration, but still result in depolarization over sufficiently long timescales.
We explicitly check this by comparing a more conventional non-repeated gate RB to the decay given by a repeated protocol of sets of 5 gates (\cref{fig:all_rb}a).
Finally, our dispersive measurement scheme's errors should only depend on the expectation value of $Z$, and thus the extra $Z$ rotation in our inverse should not affect to the protocol.
We thus expect the arguments in \cite{DirectRB} to hold for our post-selected protocol.

We first execute this protocol with no selection (other than on the final state of Bob's ancilla) and measure the success probability $P_{0_a1_b}$ of returning the system to its initial state.
In general, this curve should have two characteristic timescales corresponding to decay out of and dephasing within the dual-rail subspace.
However, as a sign of our strong noise-bias towards photon-loss errors, we find that the probability decays in a dominantly single-exponential manner (\cref{fig:all_rb}), with a decay constant of $\tau_{\text{RB}} = 1271\pm4$ gates.
We use this exponential behaviour to estimate an effective average unselected gate infidelity of $0.078\pm0.001\%$.

To accurately quantify the rate of decay, we use our measurements of Alice to find the combined probability of staying in the dual-rail subspace after each sequence,  $P_{1_a0_b\; \cup \; 0_a1_b}$. 
This probability is unaffected by dephasing within the subspace, allowing us to extract a pure leakage timescale of $\tau = 1388\pm4$ gates, corresponding to a $0.072 \pm 0.001\%$ probability of decay per gate.
It is clear that our 'raw' RB is strongly limited by this timescale, given the overlap of the two curves in \cref{fig:all_rb}.

Finally, we focus on the error-detected dataset, containing sequences with no leakage events (we ignore the minute probability of both a cavity decay and heating event occurring within these timescales).
These sequences contain trajectories which remain in the dual-rail subspace, with errors that arise from driven  dephasing, `no-jump' evolution towards the higher coherence cavity \cite{JamesDualRailPaper}, and imperfect control.
As argued above, we expect these errors to be converted to a depolarization channel under the RB protocol.
Fitting these curves with an exponential decay to an offset of 0.5, we extract time constants of $\tau = 4477\pm 30$ gates and $\tau = 4194\pm 32$ gates for the error-detected $P_{0_a1_b}$ and $P_{1_a0_b}$ data, respectively. 

Notably, the error-detected curves deviate from an exponential fit at long timescales. 
This could be due to a resurgence in ancilla-induced dephasing at timescales where the ancilla can both heat and decay, or significant probability that sequences containing photons at long timescales arise from cavity heating instead of the initial state preparation, or minor imperfections in state readout that are only visible when Alice and Bob populations are comparable.
For these reasons, we restrict our fits in the main text to the first $2250$ gates, where the decay is well-described by an exponential decay, and obtain an error-detected infidelity of $0.02 \pm 0.001\%$.
Finally, since each gate in $G_{DR}$ is constructed from one or more near-identical beamsplitter pulses, we use the fact that the average gate has $4/3$ beamsplitters to convert this gate infidelity into an error-detected beamsplitter infidelity of $0.015 \pm 0.001\%$. This factor of $4/3$ comes from the fact that $4/6$ gates in $G_{DR}$ are beamsplitter pulses and $2/6$ gates are swap operations implemented with two consecutive beamsplitters. This leads to an average of $1\times4/6 + 2\times2/6 = 4/3$ beamsplitters per gate on average in $G_{DR}$.

\section{Data normalization}
We detail in this section any schemes used to normalize data in the main text, which were used to put focus on the relative decay of fidelity as a function of time or number of gates, instead of SPAM errors.
The coherent evolution data shown in Fig. 2b of the main text was normalized with respect to a separately measured readout infidelity.
This measurement involved executing the standard protocol for preparing a single photon in Bob (described in the main text) and probing its state through selective spectroscopy of Bob's ancilla.
The resulting number-split peaks showed population at $\vert 0_b \rangle$ and $\vert 1_b \rangle$ respectively, with
no measureable population at Fock states $\vert 2_b \rangle$ and higher.
A simple linear transformation was then applied to the spectroscopy data such that the offset was set to zero, and the total population in $\vert 0_b \rangle$ and $\vert 1_b \rangle$ exactly summed to one.
The re-normalized population in $\vert 1_b \rangle$ was inferred to be the state preparation fidelity ($\sim 94\%$).
An identical linear transform was applied to the coherent evolution data.

The RB datasets (shown in Fig. 3 of the main text) were normalized in different ways.
The raw and the coupler-selected datasets were both exactly normalized to span between 1.0 (at zero gates) and 0.0 (at 8100 gates), to ignore SPAM errors.
The error-detected dataset was normalized such that it had a value of 1.0 at zero gates, and a steady-state value of 0.5 would map to a normalized value of 0.0.
This steady-state estimate of 0.5 was chosen due to a lack of high-confidence data points at very long times, to allow a visual comparison of the error-detected dataset to the raw dataset.
All unnormalized RB curves are shown in \cref{fig:all_rb}.

\section{Device Fabrication}
\label{fab_notes}
\begin{figure}[ht]
\centering
\includegraphics[width=0.95\columnwidth]{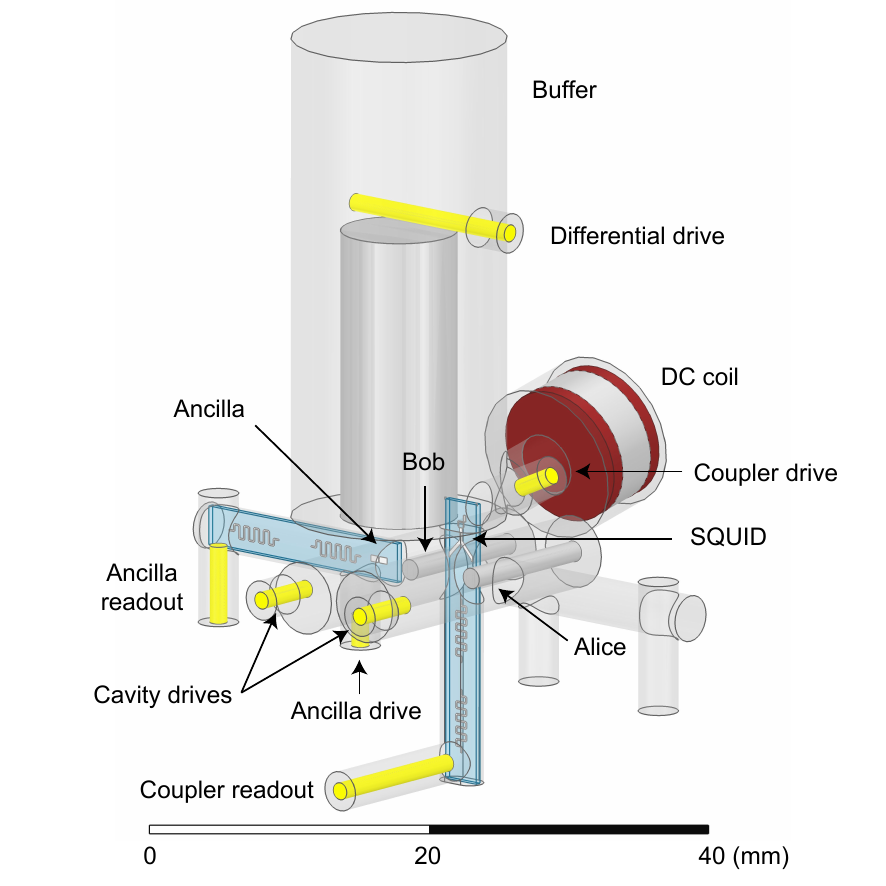}
\caption{ \textbf{Illustration of the full package.}
An x-ray view showing the relative orientations of the buffer cavity, coupler, Alice, Bob, the ancilla and the electromagnetic coil used for DC flux delivery.
The non-reflective grey space is vacuum, and the grey co-axial posts in the cavities are high-purity Aluminum. Alice, Bob and the buffer mode are the three coaxial stub cavities with designed stub lengths of $\approx$11.2mm, 10.7mm and 21.4mm, respectively.
Both the coupler and the ancilla have on-chip stripline resonators that act as Purcell filters and readouts respectively.
Yellow rods denote capacitively coupled transmission line 'pins' that are used to address each mode.
}\label[fig]{fig:buffer_package}
\end{figure}

We detail here the fabrication of the superconducting circuits used in our experiment.
The coupler and the ancilla transmon chip devices are fabricated with e-beam lithography on a 2-inch, 430 µm thick C-plane EFG sapphire substrate. E-beam writing of the Josephson junctions (designed in the `Manhattan' style with 90$^{\circ}$ cross-over), as well as the coarser patterns, is performed in Raith EBPG 5000+. Two aluminum layers are deposited via dual-angle electron-beam
evaporation in a Plassys UMS 300, and are gapped by an $\text{Al}_\text{x} \text{O}_\text{y}$ insulator layer grown in an oxidation process under 20 mbar of 85\% Ar and 15\% $\text{O}_2$ mixture for 12 minutes. For both chip devices, two adjacent stripline resonators with meandering-line geometry are used as the on-chip readout mode and Purcell filter mode (\cref{fig:buffer_package}), with the readout mode dispersively coupled to the coupler/ancilla mode, and the Purcell filter mode strongly coupled to the readout port.
The superconducting package hosting the chips is made of a single piece of high-purity $99.999\%$ ('5N') Aluminum and is chemically etched to improve surface quality.

\section{Device Parameters}
We display in \cref{coherences,Kerrs} the measured device parameters of the SQUID coupler, Alice, Bob, and Bob's transmon ancilla. 
Here, $T_2^*$ and $T_2^E$ are extracted from Ramsey and $T_2$ Echo experiments, respectively. 
The Coupler $T_1$ has been observed to drift between $40$ and $80\mu$s within and between cooldowns.
Bob's diminished coherence times as compared to Alice are likely to do the smaller detuning between Bob and the Coupler, leading to increased participation in the bare coupler mode, as well as deleterious effects of the Transmon ancilla. In particular, Bob's $T_2$ likely suffers due to the relatively high thermal population (and thus heating rate) of the ancilla. 
Cross-Kerrs between the ancilla and the distant Alice and coupler modes were not directly measured, but are assumed to be negligible. From the cross-Kerrs measured between the coupler and the storage cavities, we infer their direct linear coupling strengths to be $g_{ac}/2\pi\approx 82$ MHz, and $g_{bc}/2\pi\approx 78$ MHz.
\begin{table}[ht]
\centering
\begin{tabular}{*5c}
\toprule
 & Coupler   & Alice  & Bob & Ancilla \\
\midrule
Frequency (GHz) & 7.245 & 6.225 & 6.46 & 5.663\\ 
$T_1$ ($\mu$s) & $\sim$60 &  375 & 300 & 120\\
$T_2^*$ ($\mu$s) & 25-35 &  450 & 250 & 5.5\\ 
$T_2^E$ ($\mu$s) & 40 &  N/A & N/A & 25\\ 
$n_{th}$ & 0.02 & $<$0.03 & $<$0.03 & 0.06\\ 
\bottomrule\\
\end{tabular}
\caption{Measured frequencies, coherence times, and thermal populations for the SQUID's coupler mode, the storage modes Alice and Bob, and Bob's coupled ancilla transmon.}
\label[table]{coherences}
\end{table}

\begin{table}[ht]
\centering
\begin{tabular}{*5c}
\toprule
Kerrs & Coupler   & Alice  & Bob & Ancilla \\
\midrule
Coupler & -125~MHz & -1.7~MHz & -2.6~MHz & N/A  \\ 
Alice &  &  -4.9~KHz & -11~KHz & N/A\\
Bob &  &   & -14.6~KHz & -1.2~MHz\\ 
Ancilla &  &   &  & -180~MHz\\
\bottomrule\\
\end{tabular}

\caption{Measured self-Kerrs and cross-Kerrs for the four modes. The cross-Kerr between the ancilla and the coupler or Alice modes have not been measured, but are assumed to be negligible.}
\label[table]{Kerrs}
\end{table}

\section{Wiring Diagram}
\begin{figure*}[p!]
\centering
\includegraphics[width=0.98 \textwidth]{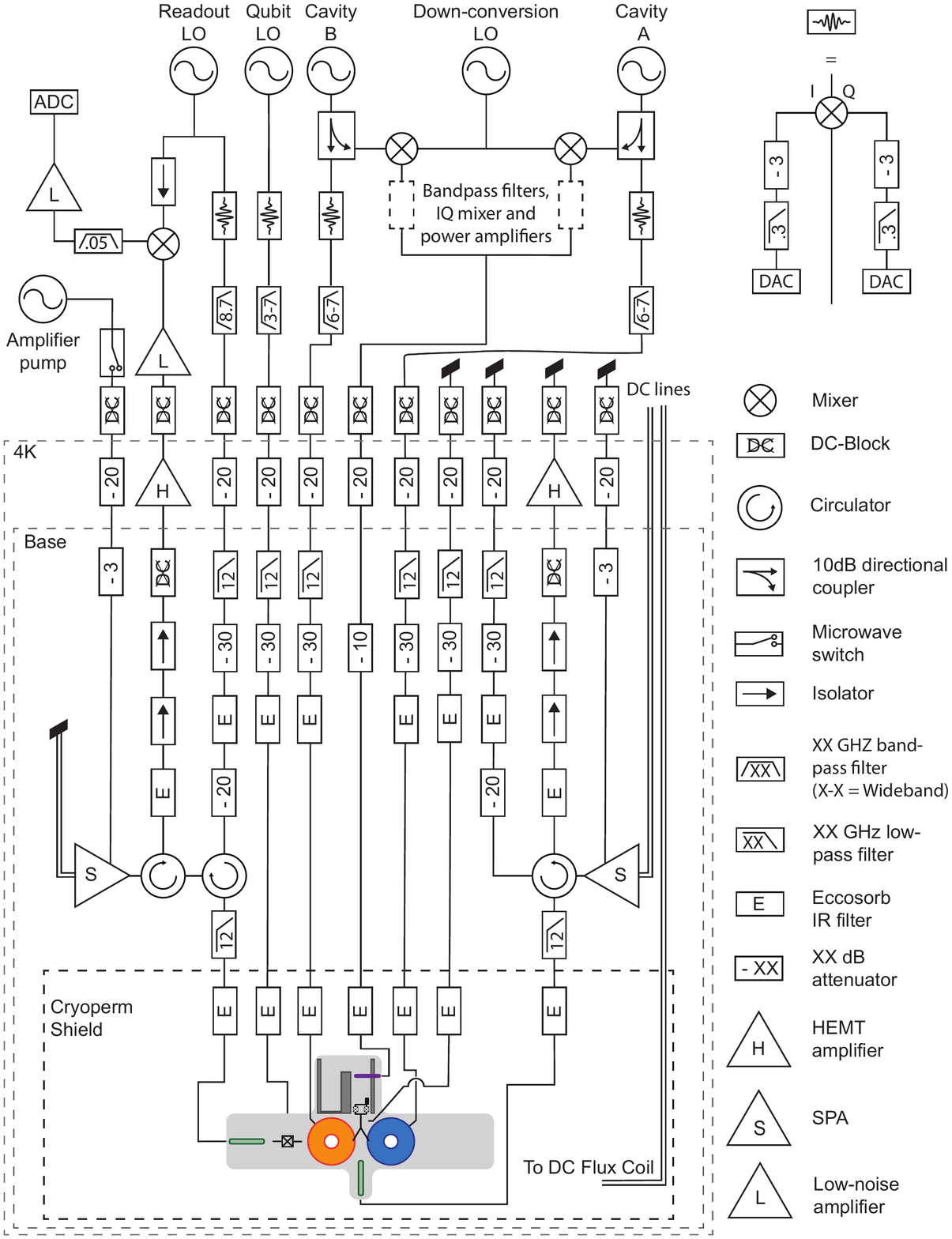}
\caption{ \textbf{Wiring Diagram}
}\label[fig]{wiring_diagram}
\end{figure*}

The schematic of the instrumentation and cryogenic setup can be seen in \cref{wiring_diagram}. The coupler and the ancilla transmon chip devices are held by chip clamps that are bolted to the superconducting Aluminum package. The package is then heat sunk to the base stage (at $\sim$20 mK) of an Oxford Instruments dilution refrigerator via an OFHC copper post. A Cryoperm can surrounds the package and provides magnetic shielding, with an inner layer of Berkeley-black-coated copper shim (not shown) acting as an IR photon absorber. seven microwave lines with different filtering schemes connect the package to the room-temperature (RT) setup: the drive lines and the readout lines for preparing and reading out the states of the coupler (common mode) and the ancilla, the drive lines for displacing Alice and Bob, and a buffer-mode drive line for applying the two RF flux drives. 

To maintain the phase stability of the beamsplitter pulse relative to the cavity drives, three local oscillators (LOs, Agilent N5183A RF signal generators) are
shared between Alice, Bob and the buffer mode, as indicated in \cref{wiring_diagram}. The down-conversion LO ($\omega_\Delta = 9.2$ GHz) mixes with the Alice LO ($\omega_a = 6.29$ GHz) and the Bob LO ($\omega_b = 6.525$ GHz), creating two drive tones at $\omega_{d_2}=2.91$GHz and $\omega_{d_1}=2.675$GHz that serve as effective LOs for the two flux drives. The Coupler, ancilla and readout drives are created by independent LO sources. The
LO tones are then mixed with waveform pulses generated by four FPGA-based quantum controllers (Innovative Integration X6-1000M). The custom-built DC flux line carries the dc current generated by a YOKOGAWA GS200 low-noise
voltage source that we use to bias the SQUID to true zero flux quantum. The DC flux line is made of copper from RT to 4K, soldered to NbTi superconducting wire from 4K to below, with the solder joint thermalized to the 4K plate to ensure that the thermal dissipation does not exceed the cooling power of the dilution refrigerator.  
Measurement of the coupler and the ancilla is done with
dispersive readout in reflection. The readout signals 
are amplified by two SNAIL parametric
amplifiers (SPA) at the base stage, and are further amplified by HEMT amplifiers at 4 K and MITEQ amplifiers at room temperature, then
down-converted to an intermediate frequency (50 MHz)
by the same readout LO, and digitized by a pair of ADCs integrated into the FPGA cards.

\end{document}